\documentclass[reprint, twocolumn, aps, amsmath, amssymb, prd, nofootinbib]{revtex4-2}
\usepackage{graphicx}	
\graphicspath{{./}{../}}  
\usepackage{amsmath}	
\usepackage{empheq}
\usepackage{xcolor}

\usepackage{multirow}
\usepackage{makecell}
\usepackage{booktabs}
\usepackage{caption}  
\usepackage{float}    

\usepackage{ragged2e}   
\usepackage{CJKutf8}
\usepackage[colorlinks = true,
            linkcolor = blue,
            urlcolor  = blue,
            citecolor = blue,
            anchorcolor = blue]{hyperref}

\newcommand{\be}{\begin{equation}}
\newcommand{\ee}{\end{equation}}
\newcommand{\bea}{\begin{eqnarray}}
\newcommand{\eea}{\end{eqnarray}}
\def\ba#1\ea{\begin{align}#1\end{align}}

\newcommand{\reffig}[1]{Figure~\ref{fig:#1}}

\newcommand{\refsec}[1]{Section~\ref{sec:#1}}

\newcommand{\reftab}[1]{Table~\ref{tab:#1}}

\def\({\left(}
\def\){\right)}
\def\[{\left[}
\def\]{\right]}
\def\<{\left<}
\def\>{\right>}

\def\d{{\rm d}}

\def\ln{{\rm ln}}

\definecolor{RedWine}{rgb}{0.743,0,0}

\definecolor{NavyBlue}{rgb}{0,0,0.9}

\usepackage[normalem]{ulem}
\definecolor{RevPink}{rgb}{1,0.08,0.58}

\begin{document}

\title
{Model-independent Constraints on Dark Energy Density from CMB-SPA, DESI DR2 BAO, and Updated Type Ia Supernova Compilations}

\author{Zhenyuan Wang (\begin{CJK*}{UTF8}{gbsn}王震远\end{CJK*})}
\email{zywang@ipac.caltech.edu}
\author{Yun Wang}
\email{wang@ipac.caltech.edu}

\affiliation{IPAC, California Institute of Technology, 1200 E. California Blvd., Pasadena, CA 91125, U.S.A.}

\begin{abstract}
Recent combinations of DESI BAO, CMB, and Type~Ia supernova data report a preference for evolving dark energy that is obtained within the $(w_0, w_a)$ parameterization and depends on the choice of SN~Ia compilation. We constrain the dark energy density ratio $X(z) \equiv \rho_{\rm DE}(z)/\rho_{\rm DE}(0)$ at five redshifts over $0 < z < 2.33$, without assuming a parametric form for the equation of state, using CMB data from SPT-3G, Planck~2018, and ACT~DR6 (hereafter CMB-SPA), DESI~DR2 BAO, and the updated Pantheon+, Union3.1, and DES-Dovekie SN~Ia compilations. For all three compilations, $X(z)$ rises above unity at low redshift, crosses unity near $z \approx 1$, and falls below unity at higher redshift. Depending on the SN~Ia compilation, the preference for a departure from a cosmological constant is $2.3$--$2.8\sigma$ with the five $X(z)$ knots, compared with $3.0$--$3.2\sigma$ in the $(w_0, w_a)$ parameterization. Because $X(z)$ is constrained locally in redshift, we can trace how the $\Lambda$CDM tension between each pair of probes propagates into its shape. The tension between BAO and the CMB in the $h r_d$--$\Omega_m$ plane reproduces this ``see-saw'' shape, whereas the $\Omega_m$ tension between SNe~Ia and the CMB only raises $X(z)$ slightly and nearly uniformly. The origin of the BAO--CMB tension can be tested by Euclid and Roman BAO measurements at $z \gtrsim 1$. We derive the distance prior $(R, \ell_A, \omega_b)$ from CMB-SPA data and show that it retains essentially all the information on $X(z)$ in the CMB-SPA power spectra without the lensing reconstruction, and that adding the lensing reconstruction only marginally increases the information on dark energy. Our work demonstrates the importance of model independence in exploring the nature of dark energy.
\end{abstract}



\maketitle


\section{Introduction}
\label{sec:intro}

Recent measurements from DESI have renewed scrutiny on whether the dark energy driving cosmic acceleration~\cite{Riess1998, Perlmutter1999} is consistent with a cosmological constant. Combining DESI DR2 Baryon Acoustic Oscillation (BAO) with cosmic microwave background (CMB) data, the preference for a deviation from a cosmological constant reaches 3.1$\sigma$ \cite{DESIDR2, DESIDR2_DE} in the $(w_0,w_a)$ parameterization \cite{Chevallier2001, Linder2003}. Adding Type Ia supernova (SN~Ia) datasets shifts this preference to between $2.8\sigma$ and $4.2\sigma$, depending on which compilation is used \cite{DESIDR2, Scolnic2022, Brout2022, Rubin2023, Abbott2024} (see \cite{Giare2025, Capozziello2026} for reviews of the preference across different dataset combinations including SDSS, Big Bang Nucleosynthesis or cosmic chronometers). Because these results are all expressed in the two-parameter form of the equation of state $w(z)$, which is linear in the scale factor, they neither locate in redshift the departure from a cosmological constant, nor show transparently how the tensions among the CMB, BAO, and SN~Ia datasets under $\Lambda$CDM could contribute to it.

Since we do not know the true nature of dark energy at present, the natural choice is to constrain the \emph{density} of dark energy, rather than its equation of state $w(z)$~\cite{WangFreese2006}, from these geometric probes (CMB, BAO, and SN~Ia). All these distance measurements depend on the Hubble parameter $H(z)$ (or its shape as a function of redshift), which is determined by the dark energy density as a function of redshift. In contrast, any assumed equation of state $w(z)$ must be integrated over redshift to obtain the dark energy density before the Hubble parameter can be predicted,
\begin{equation}
    \rho_{\rm DE}(z) = \rho_{\rm DE}(0) \exp\left[ 3\int_0^z \d z^{\prime}\frac{1+w(z^\prime)}{1+z^\prime} \right].
\end{equation}
As a result, the constraints on $w(z)$ at any given redshift depend on the entire integrated history of $w(z)$, so that a parameterized form such as $(w_0,w_a)$ imposes correlations across redshifts and can hide where the data actually prefer a departure from a cosmological constant. Truncating the expansion of $w(z)$ at first order also adds information that is not present in the data, and marginalizing over the higher-order terms instead weakens the evidence against a cosmological constant~\cite{Nesseris2026}.

In this paper, we constrain the dark energy \emph{density ratio} $X(z) \equiv \rho_{\rm DE}(z) / \rho_{\rm DE}(0)$ in a \emph{model-independent} way, at five redshift knots over $0 < z < 2.33$. The dark energy density ratio at any redshift between these knots is obtained by cubic spline interpolation. A spline interpolation assumes no global structure of $X(z)$, unlike Chebyshev polynomial expansion or Gaussian processes, and can therefore capture local features preferred by the data. This flexibility comes at the price of larger uncertainty compared to $(w_0, w_a)$ parameterization owing to a larger number of parameters \cite{ZhaiWang2019, DESIDR2_DE, WangFreese2025}. However, as we show in this paper, this price is modest given the precision of the latest CMB, BAO, and SN~Ia datasets. On the other hand, because $X(z)$ is constrained locally in redshift, one can clearly see at which redshifts the dark energy density deviates from cosmological constant $\Lambda$, while this redshift-tomographic information is usually compressed into two numbers in the $(w_0,w_a)$ parameterization. 

Several papers have performed the model-independent constraints on dark energy in the past two decades using geometric probes 
\cite{WangFreese2006, WangDai2016, ZhaiWang2019}, including recent reconstructions of the dark energy density with DESI BAO \cite{Calderon2024, DESIDR2_DE, Berti2025, WangFreese2025, Kessler2026}, and of its equation of state~\cite{WangFreese2025, Dinda2025b, SousaNeto2025}. Other non-parametric methods for reconstructing $X(z)$ or $w(z)$ include Gaussian process regression~\cite{DESIDR2_DE, Dinda2025b, deSouza2026, Akarsu2026, MukherjeeSen2025}, binned densities with a correlated prior~\cite{Wang2018}, crossing statistics~\cite{Calderon2024}, weighted function regression~\cite{GonzalezFuentes2025}, and symbolic regression~\cite{SousaNeto2025} (see also Refs.~\cite{MukherjeeSen2024, MukherjeeSen2025} for reconstructions of the expansion history). Compared to previous work, we use a cubic spline to constrain the evolution of $X(z)$ from the latest data from DESI DR2 BAO \cite{DESIDR2}, the CMB power spectra from SPT-3G~\cite{Camphuis2026}, Planck 2018~\cite{Planck18}, and ACT~DR6~\cite{Louis2025, Calabrese2025} (CMB-SPA \cite{Camphuis2026}), and three updated SN~Ia data compilations, Pantheon+ \cite{Scolnic2022, Brout2022, Hoyt2026}, Union3.1 \cite{Hoyt2026}, and DES-Dovekie \cite{Dovekie2025}. We compress the CMB-SPA likelihood into a distance prior $(R,\ell_A,\omega_b)_{\rm SPA}$ \cite{Wang2007}, and verify that it gives nearly the same $X(z)$ posterior as the full CMB-SPA power spectra, with and without CMB lensing reconstruction likelihood. We also apply flux averaging \cite{Wang2000, ZhaiWang2019} to Pantheon+ and DES-Dovekie, to validate the robustness of the results against the weak-lensing magnification and other non-Gaussian scatter of SN~Ia distances. 

Many recent studies have looked for the origin of the tensions between the CMB, BAO, and SNe~Ia within $\Lambda$CDM, either in the data themselves~\cite{Efstathiou2025, Zaldarriaga2026, Karim2026, Colgain2022, Colgain2024a, Colgain2025b, Colgain2026, Afroz2026, Dinda2025a, LopezHernandez2026} or in the calibration of individual probes~\cite{Efstathiou2025SN, Vincenzi2025, Hoyt2026, Dovekie2025, HuangCaiWang2025, Ormondroyd2026}. Others have asked whether extensions of $\Lambda$CDM other than evolving dark energy can relieve these tensions~\cite{Sailer2026, Sullivan2026, deCruzPerez2026, Chen2025, Schoneberg2026, Jhaveri2025, Weiner2026}, or have studied the connection between the tensions in $\Lambda$CDM and the dark energy parameters~\cite{Tang2025, Lee2022, Lee2025_Om, YeLin2025}. In this paper, we project the inter-probe tensions in the $\Lambda$CDM model into $X(z)$ to understand how the evolving dark energy density absorbs each of these tensions. We show that the $h r_d$--$\Omega_m$ tension between BAO and the CMB and the $\Omega_m$ tension between SNe~Ia and the CMB project into the shape of $X(z)$ in fundamentally different ways, while this distinct information is compressed into a single significance of deviation in the $(w_0,w_a)$ parameterization. This makes $X(z)$ a more informative and transparent basis for understanding the tension projection. We further show that the tension projection largely reproduces the measured $X(z)$ for CMB+BAO+SN~Ia, and that the current see-saw shape of $X(z)$ is driven mainly by the tension between BAO and the CMB. On the other hand, including the SH0ES calibration of the SN~Ia absolute magnitude \cite{Riess2022} lowers $X(z)$ at all redshifts. However, the free evolution of the dark energy density alone cannot absorb the Hubble tension, consistent with Ref.~\cite{Efstathiou2026}.

This paper is organized as follows. Section~\ref{sec:method} describes the cosmological models, the model-independent reconstruction of $X(z)$, the SN~Ia data with flux averaging and the compressed likelihood, the CMB data with the full likelihood and the CMB data compressed as distance prior, and the BAO data. Section~\ref{sec:results} presents the constraints in flat $\Lambda$CDM, flat $w_0w_a$CDM, and on the model-independent $X(z)$, and tests the robustness of $X(z)$ against flux averaging and against the compression of the CMB likelihood into the distance prior. Section~\ref{sec:disc_Om_tension} investigates the propagation of the inter-probe tension in the $h r_d$--$\Omega_m$ plane and the Hubble tension into $X(z)$. We discuss our results in the context of other studies of the inter-probe tensions and of extensions of $\Lambda$CDM, and conclude in Section~\ref{sec:discussion}. Appendix~\ref{app:fisher} derives the Fisher-matrix projection of the tension and gives the Fisher uncertainties and correlations of the $X(z)$ knots, and Appendix~\ref{app:logrp} provides the compressed SN~Ia likelihoods used in this paper.

\section{Methodology, datasets, and likelihoods}
\label{sec:method}
In this section, we introduce the method we use to constrain the dark energy density by combining CMB, BAO, and SN~Ia data. We first review the cosmological models and the distance measures in a flat universe in \refsec{distance_method}. We then introduce the details on the model-independent constraint on dark energy density ratio $X(z)$ at redshift knots in \refsec{Xz_method}, including the underlying assumptions and the best parameterization for different combinations of datasets. In \refsec{SNe_flux_avg}, we introduce the SNe~Ia compilations, flux averaging, and the compressed SN~Ia likelihood. Section~\ref{subsec:data_cmb} describes the CMB-SPA power spectra and the construction of the CMB distance prior $(R,\ell_A,\omega_b)_{\rm SPA}$, and \refsec{BAO_data} describes the DESI~DR2 BAO measurements. Finally, \refsec{likelihood} summarizes the likelihood, the priors, and the sampling of the posteriors.

\subsection{Cosmological models and distance measures}
\label{sec:distance_method}
In this paper, we assume a Friedmann--Lema\^{i}tre--Robertson--Walker metric, in which the dimensionless Hubble parameter follows
\bea 
E(z) &\equiv& H(z)/H_0 = \Big[\Omega_{bc}(1+z)^3 + \Omega_\gamma (1 + z)^4 \nonumber\\
&& + \Omega_\nu\,\frac{\rho_\nu(z)}{\rho_\nu(0)} + \Omega_k(1+z)^2 + \Omega_{\rm DE}X(z)\Big]^{1/2}.
\label{eq:Ez_general}
\eea
Here $H(z)$ and $H_0$ are the Hubble parameter and the Hubble constant. $\Omega_{bc}$, $\Omega_\gamma$, $\Omega_\nu$, $\Omega_k$, and $\Omega_{\rm DE}$ are the baryon plus cold dark matter, photon, neutrino, curvature, and dark energy density fractions at the present day. The neutrino term includes one massive species with $m_\nu = 0.06$\,eV and $N_{\rm eff} = 3.044$. Its density $\rho_\nu(z)$ evolves from radiation-like to matter-like and is computed numerically. At low redshift the massive neutrinos behave as matter, and we define $\Omega_m \equiv \Omega_{bc} + \Omega_\nu$. These fractions satisfy
\bea
\Omega_{bc} + \Omega_\gamma + \Omega_\nu + \Omega_k + \Omega_{\rm DE} = 1.
\eea
We consider three models for $X(z)$:
\begin{widetext}
\bea
X(z) \equiv \frac{\rho_\mathrm{DE}(z)}{\rho_\mathrm{DE}(0)}
= 
\begin{cases}
1 , & \Lambda \\
(1+z)^{3(1+w_0+w_a)} \exp\left(-\frac{3w_a z}{1+z}\right), & w_0w_a \\
\text{spline through } \{X_k\}, & \text{model-independent interpolation}
\end{cases}
\eea
\end{widetext}
Here $(w_0, w_a)$ is the linear parameterization of the dark energy equation of state \cite{Chevallier2001, Linder2003}: 
\bea
w(z) = w_0 + w_a \frac{z}{1 + z}.
\label{eq:w0wa}
\eea
In the model-independent approach, $\{X_k\}$ are the values of \emph{dark energy density ratio} $X(z)$ at a set of redshift knots. The evolution at arbitrary redshift is calculated through the cubic spline interpolation between these knots. We provide more details in \refsec{Xz_method}.

Throughout the paper, we assume the universe is spatially flat, i.e. $\Omega_k = 0$. The comoving distance then follows
\bea
r(z) = \frac{c}{H_0} \int_0^z \frac{\d z^\prime}{E(z^\prime)}.
\eea
The luminosity distance measured by SN~Ia is converted from comoving distance by
\bea
d_L(z) = (1 + z) r(z).
\eea
We describe how each dataset constrains these quantities in the following subsections.

\subsection{Model-independent constraints on dark energy density}
\label{sec:Xz_method}
The model-independent approach constrains the dark energy density ratio $X(z)$ directly at a set of redshift knots. This allows us to recover the evolution of the dark energy density preferred by the data without imposing a parametric form for $w(z)$. Based on the redshift coverage of the SN~Ia compilations and DESI~DR2 BAO, we place five free knots at
\bea
\{z_k\} = \{1/3,\ 2/3,\ 1,\ 4/3,\ 2.33\},
\eea
in addition to the fixed point $X(z_0 = 0) = 1$. This was shown to be the optimal choice in \cite{WangFreese2025}.
We assume that $X(z)$ varies smoothly on scales comparable to the knot spacing, and obtain $X(z)$ at any redshift by cubic-spline interpolation through the knots,
\bea
X(z) =
\begin{cases}
\displaystyle\sum_{k} S_k(z)\, X(z_k), & z \leq 2.33, \\[6pt]
X(2.33), & z > 2.33,
\end{cases}
\label{eq:Xz_interpolation}
\eea
where the spline basis functions $S_k(z)$ depend only on the knot redshifts and the boundary conditions, $X'(2.33) = 0$ at the high-redshift end, so that $X(z)$ is linear in the knot values. Since neither the BAO nor the SN~Ia data extend beyond $z = 2.33$, we have no direct constraint on $X(z)$ at higher redshift. We thus assume $X(z) = X(2.33)$, i.e. $w = -1$ at $z > 2.33$, which joins the spline smoothly because $X'(2.33) = 0$. This choice does not affect the BAO or SN~Ia likelihoods, which only need theoretical predictions at $z \leq 2.33$. It does enter the CMB through the distance to the last-scattering surface. Nevertheless, we expect this assumption to have a subdominant effect because dark energy accounts for only a few percent of the total energy budget at $z \geq 2.33$, and less at higher redshift. Any small change to this small fraction should be negligible given the current constraining power of the data. When converting $X(z)$ into the equation of state $w(z)$ required by \texttt{CAMB} to compute the CMB power spectra (Section~\ref{subsec:data_cmb}), we reject the steps in which $w(z)$ leaves $[-10, 10]$ at any redshift, to avoid the numerical instabilities caused by rapid oscillations of $X(z)$ between the knots. We verified that this wide boundary on $w(z)$ does not impact the posterior of $X(z)$.

For a spatially flat universe, Equation~\eqref{eq:Ez_general} becomes
\bea
E^2(z) &=& \Omega_{bc}(1+z)^3 + \Omega_\gamma(1+z)^4 + \Omega_\nu\,\frac{\rho_\nu(z)}{\rho_\nu(0)} \nonumber\\
&& + \left(1 - \Omega_{bc} - \Omega_\gamma - \Omega_\nu\right) X(z).
\label{eq:Ez_full}
\eea
The radiation and massive-neutrino terms are needed to extend $E(z)$ to the high redshifts relevant for the CMB. At $z \lesssim 2.33$, the massive neutrinos behave as matter and the radiation term is negligible, so Equation~\eqref{eq:Ez_full} reduces to
\bea
E^2(z) &\simeq& \Omega_m(1+z)^3 + (1 - \Omega_m) X(z), \nonumber\\
\Omega_m &\equiv& \Omega_{bc} + \Omega_\nu.
\label{eq:Ez_lowz}
\eea
We compute the full Equation~\eqref{eq:Ez_full} with \texttt{CAMB} whenever CMB data are used and use Equation~\eqref{eq:Ez_lowz} otherwise.

Note that we reparameterize $X(z)$ and $h$ when combining only the CMB and BAO datasets. Without SNe~Ia, the CMB fixes $\omega_m$ and the sound horizon $r_d$, and the BAO calibrated by $r_d$ then measures the absolute expansion rate $H(z)$, the comoving distance $r(z)$, or their combination at $z \gtrsim 0.295$. Since $H^2(z) /[100\,{\rm km\,s^{-1}\,Mpc^{-1}}]^2 \simeq \omega_m(1+z)^3 + \omega_\gamma(1+z)^4 + \omega_{\rm DE}(z)$, CMB and BAO together determine the physical dark energy density at the knots, which is defined as 
\bea
\omega_{\rm DE}(z_k) \equiv \Omega_{\rm DE} h^2 X(z_k). 
\eea
They constrain its present-day value, $\omega_{\rm DE}(0) = h^2 - \omega_m - \omega_\gamma$, only weakly because no absolute distance measurement anchors $h$ below $z \approx 0.295$, so $h$ is fixed only through the extrapolation of the spline to $z = 0$. The ratio $X(z_k) = \omega_{\rm DE}(z_k)/\omega_{\rm DE}(0)$ therefore has a poorly constrained denominator, and sampling $X(z_k)$ directly leads to strongly non-Gaussian, prior-dominated posteriors and poorly converged chains. In other words, the overall amplitudes of $X(z_k)$ knots are degenerate with $h$. For CMB+BAO we instead sample $\omega_{\rm DE}(z_k)$ and derive $X(z_k)$ afterwards. The small-$\omega_{\rm DE}(0)$ tail of the posterior makes the derived $X(z_k)$ skewed toward large values (Section~\ref{subsec:res_Xz}). Note that flat priors on $\omega_{\rm DE}(z_k)$ do not correspond to flat priors on $X(z_k)$, which should be kept in mind when comparing with the results that include SNe~Ia. When SNe~Ia are included, they constrain the shape of $E(z)^2 \equiv (H(z)/H_0)^2 \simeq \Omega_m(1+z)^3 + (1-\Omega_m)X(z)$ down to $z \approx 0.01$. Therefore the combination of CMB, BAO, and SNe~Ia is able to constrain the overall amplitude of $X(z)$. We then sample $X(z_k)$ directly.

The same method could also be applied to measure the dark energy equation of state $w(z)$ at redshift knots. However, $w(z)$ reconstructed with a cubic spline~\cite{WangFreese2025} or other non-parametric methods~\cite{Gu2025, Ye2025, Yang2025, Ghedini2026, WangLiFan2026} often shows wiggles. Because the distance $r(z)$ depends on $w(z)$ through two-fold integrals~\cite{WangFreese2006}, such wiggles are largely smoothed out in the observables. Such wiggles are therefore allowed by the data but may be spurious~\cite{Akarsu2023}. On the other hand, a reconstruction of $w(z)$ guarantees a positive dark energy density, whereas a reconstruction of $\rho_{\rm DE}(z)$ does not and requires a lower prior bound at zero (Table~\ref{tab:priors}). In our analysis this bound has negligible effect. Whether we sample $\omega_{\rm DE}(z_k)$ for CMB+BAO or $X(z_k)$ when SNe~Ia are included, $\omega_{\rm DE} = 0$ and $X=0$ lie outside the 99.7\% interval of every knot.

\subsection{SN~Ia: flux averaging and compressed likelihood}
\label{sec:SNe_flux_avg}
SNe~Ia measure the distance modulus, which is converted from the luminosity distance by
\bea
\mu \equiv  m - M_B = 5 \log_{10} \[\frac{d_L(z)}{\rm Mpc}\] + 25,
\eea
where $m$ and $M_B$ are the apparent and absolute magnitudes of each supernova. Since $d_L$ is inversely proportional to $H_0$, the absolute magnitude $M_B$ and $H_0$ enter the distance modulus only through the combination 
\be
\mathcal{M} = M_B + 5 \log_{10} \left(\frac{c H_0^{-1}}{{\text{Mpc}}}\right), 
\label{eq:cal-M}
\ee
and are therefore completely degenerate. In practice, both $M_B$ and $H_0$ are analytically marginalized over in the likelihood, and only the shape of the distance--redshift relation is used to constrain the cosmological parameters.

In this paper, we use three updated SN~Ia datasets, Pantheon+~\cite{Scolnic2022, Brout2022}{\footnote{\url{https://github.com/PantheonPlusSH0ES/DataRelease}}}, 
DES-Dovekie \cite{Dovekie2025}\footnote{\url{https://github.com/des-science/DES-SN5YR}},
and Union3.1 \cite{Hoyt2026}{\footnote{\url{https://github.com/rubind/union3_release}}}.
For Pantheon+, we use the distance moduli with the low-redshift bias corrections revised in Appendix~F of {Ref.~}\cite{Hoyt2026}\footnote{\url{https://github.com/tjhoyt/panplus_CorrectedLowZ}}, together with the original Pantheon+ covariance matrix. We use the complete covariance matrix, which includes both statistical and systematic uncertainties, to perform the Bayesian inference. Pantheon+ covers $0.01 < z < 2.26$, DES-Dovekie covers $0.025 < z < 1.14$, and Union3.1 provides 22 pre-binned distance moduli over $0.05 < z < 2.26$.\footnote{We use the UNITY1.7 standardization, as in the main analysis of {Ref.~}\cite{Hoyt2026}. With the updated UNITY1.8~{\cite{Rubin2026}}, the $w_0w_a$CDM preference from CMB+BAO+Union3.1 changes by $0.1\sigma$, and $X(z)$ by at most $0.3\sigma$ (at $z=2/3$).}

To make the inference more efficient, we adopt the compressed SN~Ia likelihood of Paper~II~\cite{Wang2026SNcompression}, which compresses $\mathcal{O}(10^3)$ distance moduli and their covariance matrix into a Gaussian likelihood on 11 $\log r_p$ knots\footnote{We use the natural logarithm base for $\log r_p$.} placed at $z = 0.05, 0.15, 0.3, 0.4, 0.5, 0.6, 0.7, 0.8, 1.0, 1.6$, and $2.27$ for Pantheon+ and Union3.1, and at $z = 0.05, 0.15, 0.3, 0.4, 0.5, 0.6, 0.7, 0.8, 0.9, 1.0$, and $1.15$ for DES-Dovekie. Paper~II demonstrates that this compression preserves the cosmological information of the original distance--redshift measurements, accurately reproducing the full distance-modulus posteriors, while reducing the downstream MCMC cost by up to two orders of magnitude. Here $r_p$ is the rescaled comoving distance, and $\mathcal{W}(z)$ is an empirical exponent chosen to keep $r_p(z) \approx 1$ across $0 < z < 3$~\cite{Wang2026SNcompression}. This generalizes the constant exponent $\mathcal{W}=0.41$ of Refs.~\cite{Wang2009, ZhaiWang2019}. The rescaled distance and rescaling exponent are defined as
\bea
\label{eq:rp}
r_p(z) &\equiv& [H_0 r(z)/(cz)]\,(1+z)^{\mathcal{W}(z)},
\\
{\mathcal W}(z) &=& 0.223 + 0.186\,\frac{z}{1+z} + 0.245\,\(\frac{z}{1+z}\)^2.
\eea
 The explicit $H_0$ cancels the $c/H_0$ in $r(z)$, so $r_p$ depends only on $E(z)$ and encodes the shape of the expansion history alone. Furthermore, the combination $\mathcal{M}$ (see Equation~\eqref{eq:cal-M})
 has already been analytically marginalized over in the Gaussian covariance of 11 $\log r_p$. We denote the compressed data vector of the 11 $\log r_p$ values and its covariance by $(\boldsymbol y, \boldsymbol \Sigma_y)$ and {list them} in \reftab{logrp}. With the theoretical prediction of $y$, the SN~Ia likelihood reads
\bea
\log \mathcal{L}_{\text{SN}} = -\frac12 (\boldsymbol y - \boldsymbol y^{\rm th})^\intercal \boldsymbol \Sigma_y^{-1} (\boldsymbol y - \boldsymbol y^{\rm th}) + \text{const.}
\eea

As a robustness test, we also apply \emph{flux averaging} to suppress the systematic uncertainties due to weak lensing magnification. At $z \gtrsim 1$, weak gravitational lensing introduces a skewed, non-Gaussian scatter in magnitude space \cite{Wang1999, Wang2002, Wang2005}. Averaging in magnitude space is therefore biased by Jensen's inequality ($\langle \log F \rangle \neq \log \langle F \rangle$). Averaging in flux space instead removes this bias, since flux is conserved by lensing \cite{Wang2000, WangMukherjee2004, ZhaiWang2019}. Flux averaging contains four steps at a fixed cosmology,
\begin{enumerate}
\item Convert each distance modulus to ``flux'',
      $F_i = 10^{-0.4(\mu_{i} - 25)}$.
\item Form the model-corrected luminosity,
      $\mathcal{L}_i = F_i\, d_L^2(z_i)$.
\item Average within each redshift bin $b$,
      $\bar{\mathcal{L}}_b = N_b^{-1}\sum_{i \in b}\mathcal{L}_i$
      at $\bar{z}_b = N_b^{-1}\sum_{i \in b} z_i$.
\item Convert back to a flux-averaged distance modulus,
      $\bar\mu_b = 25 - 2.5\log_{10}[\bar{\mathcal{L}}_b/d_L^2(\bar z_b)]$ with bin covariance $\bar{\mathbf{C}}_{\mu} = \mathbf{P}\,\mathbf{C}_\mu\,\mathbf{P}^\intercal$, $P_{bi} = \mathcal{L}_i/(N_b \bar{\mathcal{L}}_b)$. 
\end{enumerate}
Note that flux averaging has to be applied to the full distance modulus data of Pantheon+ and DES-Dovekie dynamically with the sampled cosmology at each MCMC step, since the rescaling of flux requires a background cosmology. The Union3.1 compilation is pre-binned at the catalog level, so we cannot apply flux averaging there. When flux averaging is applied, the likelihood from SN~Ia is 
\bea
-2\log \mathcal{L}_{\text{SN}} &=& (\boldsymbol{\bar\mu} - \boldsymbol{\mu}^{\rm th})^{\intercal} 
{\bar{\mathbf{C}}_{\mu}}^{-1} 
(\boldsymbol{\bar\mu} - \boldsymbol{\mu}^{\rm th}) 
\nonumber\\
&& - 
\frac{\left[ (\boldsymbol{\bar\mu} - \boldsymbol{\mu}^{\rm th})^\intercal \bar{\mathbf{C}}_{\mu}^{-1} \mathbf{1} \right]^2}
{\mathbf{1}^\intercal \bar{\mathbf{C}}_{\mu}^{-1} \mathbf{1}}
\nonumber\\
&& + \log \det \bar{\mathbf{C}}_{\mu} + \log\left(\mathbf{1}^\intercal \bar{\mathbf{C}}_{\mu}^{-1} \mathbf{1}\right) + \text{const.}
\nonumber\\
\label{eq:chi2_marg}
\eea
The second term arises from analytically marginalizing over the combination of constant offsets $\mathcal{M} = M_B + 5 \log_{10} (c H_0^{-1}{\text{Mpc}}^{-1})$ with a flat prior, which also produces the fourth term. The third and fourth terms come from the fact that the covariance of the flux-averaged distance moduli depends on the sampled cosmology. This is in contrast to the case without flux averaging, where the covariance matrix is cosmology-independent. We divide $0.01 \leq z \leq z_{\rm max}$ into 40 bins of equal width in redshift, where $z_{\rm max}$ is the highest redshift of each compilation, and discard the empty bins, which leaves 30 bins for Pantheon+ and 39 bins for DES-Dovekie. The bin membership is fixed by the observed redshifts and does not change with the sampled cosmology. For DES-Dovekie, we additionally apply a quality cut $\sigma_\mu < 0.5$, retaining 1726 of the 1820 SNe~Ia. Because flux averaging weights all SNe~Ia within a bin equally, a few objects with large distance uncertainties can dominate the value and uncertainty of the binned distance modulus. This is a concern for DES-Dovekie, which is photometrically classified and has a much broader distribution of $\sigma_\mu$ than the spectroscopically confirmed Pantheon+ sample, for which no such cut is needed. Since the cut depends on $\sigma_\mu$ rather than on $\mu$, it does not bias the distance moduli. We also validate that removing these SNe~Ia barely changes the posterior of cosmological parameters in the analysis without flux averaging.

\subsection{CMB-SPA: full likelihood and distance prior}
\label{subsec:data_cmb}
The power spectra of CMB anisotropies contain a wealth of information that is complementary to BAO and SN~Ia distance measurements. We use the CMB information in two forms: the temperature and polarization power spectra of SPT-3G, Planck 2018, and ACT~DR6~{\cite{Louis2025, Calabrese2025}} (``CMB-SPA'' hereafter), with or without the CMB lensing likelihood~\cite{Camphuis2026}\footnote{SPT-3G D1: \url{https://github.com/Lbalkenhol/candl} and \url{https://github.com/SouthPoleTelescope/spt_candl_data}; ACT~DR6: \url{https://github.com/ACTCollaboration/DR6-ACT-lite}; ACT~DR6, Planck, and SPT-3G CMB lensing: \url{https://github.com/qujia7/spt_act_likelihood}; Planck 2018 low-$\ell$ TT and SRoll2 low-$\ell$ EE as distributed with \texttt{Cobaya}.}, and the compressed CMB distance prior. 

The CMB distance prior compresses the CMB information into three numbers, the shift parameter $R$, the acoustic angular scale $\ell_A$, and the physical baryon density $\omega_b$~\cite{Wang2007, WangDai2016, LemosLewis2023}, and marginalizes over the physics of the integrated Sachs--Wolfe (ISW) effect and CMB lensing~\cite{LemosLewis2023}. These parameters encode the late-time universe only through the geometric projection to the last-scattering surface, and are insensitive to late-time perturbation effects. Their definitions are
\bea
R &\equiv& \sqrt{\Omega_m H_0^2}\, r(z_*)/c,
\\
\ell_A &\equiv& \pi\, r(z_*) / r_s(z_*),
\\
\omega_b &\equiv& \Omega_b h^2,
\eea
where the sound horizon at the recombination redshift $z_*$ is
\bea
r_s(z_*) = \frac{c}{H_0} \int_{0}^{a_*} \frac{da'}{\sqrt{3(1 + \bar{R}_b\,a')\,{a'}^4\,E^2(a')}},
\eea
with $\bar{R}_b \equiv 3\omega_b/(4\omega_\gamma)$, where the photon density $\omega_\gamma$ is fixed by the CMB temperature $T_{\rm CMB} = 2.7255$\,K \cite{Fixsen2009} (the default value in \texttt{CAMB} and \texttt{Cobaya}).

We compress the CMB distance prior from the flat $\Lambda$CDM chain using CMB-SPA power spectra including CMB lensing, with the priors of Table~\ref{tab:priors}, and denote it by $(R,\ell_A,\omega_b)_{\rm SPA}$. We compute $(R,\ell_A,\omega_b)$ with \texttt{CAMB} \cite{Lewis2000} for the posterior samples of the chains, and take their mean and covariance. Following Ref.~\cite{CMBComp}, we fix the recombination redshift to $z_* = 1090$, both in the compression and in the likelihood. For CMB distance priors that include non-zero curvature and varying neutrino mass, see {\cite{Wang2007, WangDai2016, ZhaiPark2020, CMBComp}}. Our baseline distance prior has the mean values and covariance matrix
\be
\begin{aligned}
\begin{pmatrix} R \\ \ell_A \\ \omega_b \end{pmatrix} &= \begin{pmatrix} 1.750022 \\ 301.8407 \\ 0.0224251 \end{pmatrix}, 
\\
C &= 10^{-9} \begin{pmatrix} 9193.4 & -7243.1 & -112.86 \\ -7243.1 & 4580492 & 1712.4 \\ -112.86 & 1712.4 & 8.9879 \end{pmatrix}.
\end{aligned}
\label{eq:cmbspa_prior}
\ee
We have validated that the CMB distance prior constructed from the CMB power spectra without lensing reconstruction likelihood barely changes the posterior of downstream cosmological analyses in this paper.

Because $(R,\ell_A,\omega_b)_{\rm SPA}$ is compressed under flat $\Lambda$CDM, its use for evolving dark energy relies on the distance prior being insensitive to late-time physics. We therefore use both the full power spectra and the distance prior in the model-independent dark energy constraint, to verify that our result is not sensitive to the dark energy model assumed in the compression.  The two agree to within {$0.11\sigma$} on all $X(z)$ knots (Section~\ref{subsec:res_Xz}). Using the full CMB power spectra requires modifying \texttt{CAMB} so that the spline-interpolated $X(z)$ enters the theory prediction self-consistently. We then use the \texttt{CAMB} theory modules to compute the model spectra from the cosmological parameters. The non-linear corrections to the lensing potential are computed with HMcode-2020 \cite{Mead2021}, which takes an effective $(w_0, w_a)$ of the dark energy model. For a spline $X(z)$, the local expansion of $w(a)$ at $a = 1$ often gives $w_0 + w_a \gtrsim 0$, for which HMcode fails to converge. We therefore set the effective values used by HMcode to $(w_0, w_a) = (-1, 0)$, while the background, the linear perturbations, and the linear matter power spectrum passed to HMcode use the full $X(z)$.

When only background-level observables are used (the compressed CMB distances, BAO distance ratios, and SN~Ia $\log r_p$), dark energy perturbations do not enter, and the parameterized post-Friedmann (PPF) prescription \cite{HuSawicki2007, Fang2008} serves only to regularize the phantom crossing. When the full CMB power spectra are used, the PPF prescription additionally fixes the dark energy perturbations, for which we adopt a sound speed $c_s^2 = 1$ and vanishing anisotropic stress. We then convert the $\{X(z_k)\}$ knots to a dense $w(a)$ grid using cubic spline interpolation and pass it to \texttt{CAMB.DarkEnergyPPF.set\_w\_a\_table}.

\subsection{DESI DR2 BAO}
\label{sec:BAO_data}
We use the baseline DESI DR2 BAO measurements (Table~IV of {Ref.~}\cite{DESIDR2}) of the transverse comoving distance $D_M(z)/r_d$ and the Hubble distance $D_H(z)/r_d$ at six effective redshifts spanning $0.51 \leq z \leq 2.33$ and the BGS measurement of the isotropic distance $D_V(z)/r_d$ at $z = 0.295$, as implemented in the \texttt{bao.desi\_dr2} likelihood of \texttt{Cobaya}. In a spatially flat universe, the corresponding distances are
\bea
D_M(z) &=& \int_0^z \frac{c\,\d z^{\prime}}{H(z^{\prime})} = r(z),
\\
D_H(z) &=& \frac{c}{H(z)},
\\
D_V(z) &\equiv & \[z D_M^2(z) D_H(z)\]^{1/3}.
\eea
Note that BAO measures these distances only in units of the sound horizon at the drag epoch $r_d \equiv r_s(z_d)$, which itself cannot be determined by BAO alone. When we include the CMB likelihood in the analysis, we compute $r_d$ with \texttt{CAMB} from $\omega_b$, $\omega_c$, $N_{\rm eff}$, and $m_\nu$. When we use BAO alone to constrain the $\Lambda$CDM model, we only sample $\Omega_m$ and $h r_d$.

The BAO likelihood is evaluated as
\begin{equation}
    -2 \log \mathcal{L}_{\rm BAO} = \Delta \mathbf{d}^{\intercal}\, \mathbf{C}_{\rm BAO}^{-1}\, \Delta \mathbf{d},
    \label{eq:chi2_bao}
\end{equation}
where $\Delta \mathbf{d}$ is the 13-element residual vector containing $D_M/r_d$ and $D_H/r_d$ at each of the six higher effective redshifts and $D_V/r_d$ from BGS at $z=0.295$, and $\mathbf{C}_{\rm BAO}$ is the block-diagonal covariance over the seven redshifts, including the correlation between $D_M/r_d$ and $D_H/r_d$ at each redshift above $z=0.295$.

\subsection{Likelihood and parameter inference}
\label{sec:likelihood}
We use the cosmological inference code \texttt{Cobaya}~\cite{Cobaya2021}\footnote{\url{https://github.com/CobayaSampler/cobaya}} to sample the posteriors in the parameter space with the Metropolis--Hastings Markov chain Monte Carlo (MCMC) sampling. The priors on all sampled parameters are listed in Table~\ref{tab:priors}. When the CMB distance prior is used, $\tau$, $A_s$, and $n_s$ do not affect the background observables. With the full CMB-SPA likelihood, the Gaussian priors on $A_{\rm ACT}$ and $T_{\rm cal}$ are the external calibration priors of the ACT~DR6 and SPT-3G~D1 likelihoods. We remove the internal $\tau$ prior of the SPT-3G likelihood, so that $\tau$ is constrained only by the Planck low-$\ell$ EE (SRoll2) likelihood within its uniform prior\footnote{SPT-3G D1 \cite{Camphuis2026} uses a Gaussian prior on $\tau$ instead of using the Planck low-$\ell$ EE (SRoll2) likelihood.}. Combining all the likelihoods above, the total likelihood is
\bea
\log \mathcal{L} = \log \mathcal{L}_{\rm CMB} + \log \mathcal{L}_{\rm BAO} + \log \mathcal{L}_{\rm SN}.
\eea
All posterior distributions are sampled with 4 independent chains. Convergence is reached when all parameters satisfy the Gelman--Rubin statistic $\hat R - 1 < 0.005$ \cite{GelmanRubin1992}.

Throughout Sections~\ref{sec:results} and \ref{sec:disc_Om_tension}, we label the data as follows. ``CMB-SPA'' denotes the full power-spectrum likelihood, including the lensing reconstruction unless stated otherwise. $(R,\ell_A,\omega_b)_{\rm SPA}$ denotes the distance prior compressed from the $\Lambda$CDM chain of CMB-SPA. ``DESI~DR2'' denotes the DESI DR2 likelihood, and each SN~Ia compilation is denoted by its name, with ``(flux avg.)'' denoting flux averaging (for Pantheon+ or DES-Dovekie).

\begin{table*}[t]
\centering
\caption{Priors on the sampled parameters. $\mathcal{U}[a,b]$ denotes a uniform prior and $\mathcal{N}(\mu,\sigma)$ a Gaussian prior. The parameters below the middle rule are sampled only with the CMB-SPA power spectra likelihood (Section~\ref{subsec:data_cmb}). We assume the universe is flat and omit the word ``flat'' in front of model names for brevity.}
\label{tab:priors}
\small
\renewcommand{\arraystretch}{1.15}
\begin{tabular}{@{}llll@{}}
\toprule
Parameter & Prior & Model & Data \\
\midrule
$\omega_b \equiv \Omega_b h^2$ & $\mathcal{U}[0.005,\,0.1]$ & all & with CMB \\
$\omega_c \equiv \Omega_c h^2$ & $\mathcal{U}[0.001,\,0.99]$ & all & with CMB \\
$H_0$ [km\,s$^{-1}$\,Mpc$^{-1}$] & $\mathcal{U}[40,\,150]$ & all & with CMB \\
$w_0$ & $\mathcal{U}[-3,\,1]$ & $w_0w_a$CDM & all \\
$w_a$ & $\mathcal{U}[-5,\,3]$ & $w_0w_a$CDM & all \\
$X(z_k)$ & $\mathcal{U}[0,\,5]$ & $X(z)$CDM & CMB+BAO+SN~Ia \\
$\omega_{\rm DE}(z_k)$ & $\mathcal{U}[0,\,2]$ & $X(z)$CDM & CMB+BAO \\
{$\Omega_m$} & $\mathcal{U}[0.1,\,0.6]$ & $\Lambda$CDM & BAO or SN~Ia alone \\
{$h r_d$ [Mpc]} & $\mathcal{U}[80,\,130]$ & {$\Lambda$CDM} & {BAO alone} \\
\midrule
$\tau$ & $\mathcal{U}[0.01,\,0.1]$ & $\Lambda$CDM, $X(z)$CDM & \,\, CMB-SPA \\
$\ln(10^{10}A_s)$ & $\mathcal{U}[1.61,\,3.91]$ & $\Lambda$CDM, $X(z)$CDM & \,\, CMB-SPA \\
$n_s$ & $\mathcal{U}[0.8,\,1.2]$ & $\Lambda$CDM, $X(z)$CDM & \,\, CMB-SPA \\
$A_{\rm ACT}$ ($= A_{\rm Planck}$) & $\mathcal{N}(1,\,0.003)$ & $\Lambda$CDM, $X(z)$CDM & \,\, CMB-SPA \\
$P_{\rm ACT}$ & $\mathcal{U}[0.9,\,1.1]$ & $\Lambda$CDM, $X(z)$CDM & \,\, CMB-SPA \\
$E_{\rm cal}$ (SPT-3G) & $\mathcal{U}[0.9,\,1.1]$ & $\Lambda$CDM, $X(z)$CDM & \,\, CMB-SPA \\
$T_{\rm cal}$ (SPT-3G) & $\mathcal{N}(1,\,0.0036)$ & $\Lambda$CDM, $X(z)$CDM & \,\, CMB-SPA \\
\bottomrule
\end{tabular}
\end{table*}

\section{Cosmological constraints}
\label{sec:results}
In this section we present the cosmological constraints from the datasets described in Section~\ref{sec:method}, proceeding from the simplest model (flat $\Lambda$CDM) to the most flexible one (flat $X(z)$CDM). We first present the $\Omega_m$ posterior with each probe separately in flat $\Lambda$CDM in Section~\ref{subsec:res_flatLCDM}. This examines the level of consistency among CMB-SPA, DESI~DR2 BAO, and the three SN~Ia compilations before any evolution of dark energy is allowed, and sets the baseline for the rest of the paper. We then fit flat $w_0w_a$CDM to the combinations CMB+BAO, CMB+SN~Ia, and CMB+BAO+SN~Ia (Section~\ref{subsec:res_flatCPL}), to connect with the preference for evolving dark energy reported in the DESI official paper \cite{DESIDR2} with the latest data. Finally, we reconstruct the dark energy density ratio $X(z)$ at five redshifts with cubic spline interpolation (Section~\ref{subsec:res_Xz}), and test its robustness against flux averaging. We further test the dependence of the CMB distance prior on late-time physics by comparing the distance prior (compressed from $\Lambda$CDM chains) with the full CMB-SPA power spectra with and without lensing. Throughout, we quote posterior means with $68\%$ C.L. ranges. We quantify the preference for a departure from $\Lambda$CDM with $\Delta\chi^2_{\rm MAP}$ and Wilks' theorem (Section~\ref{subsec:res_flatCPL} and Section~\ref{subsec:res_Xz}).

\subsection{$\Lambda$CDM constraints from individual probes}
\label{subsec:res_flatLCDM}

\begin{figure*}[t]
    \centering
    \includegraphics[width=0.498\linewidth]{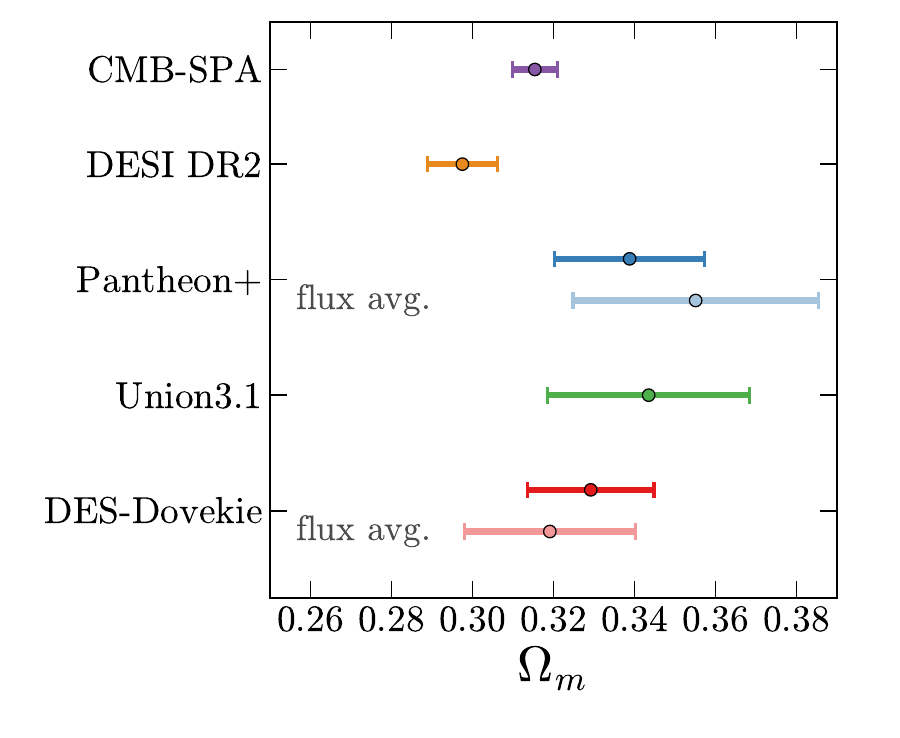}\hfill
    \includegraphics[width=0.498\linewidth]{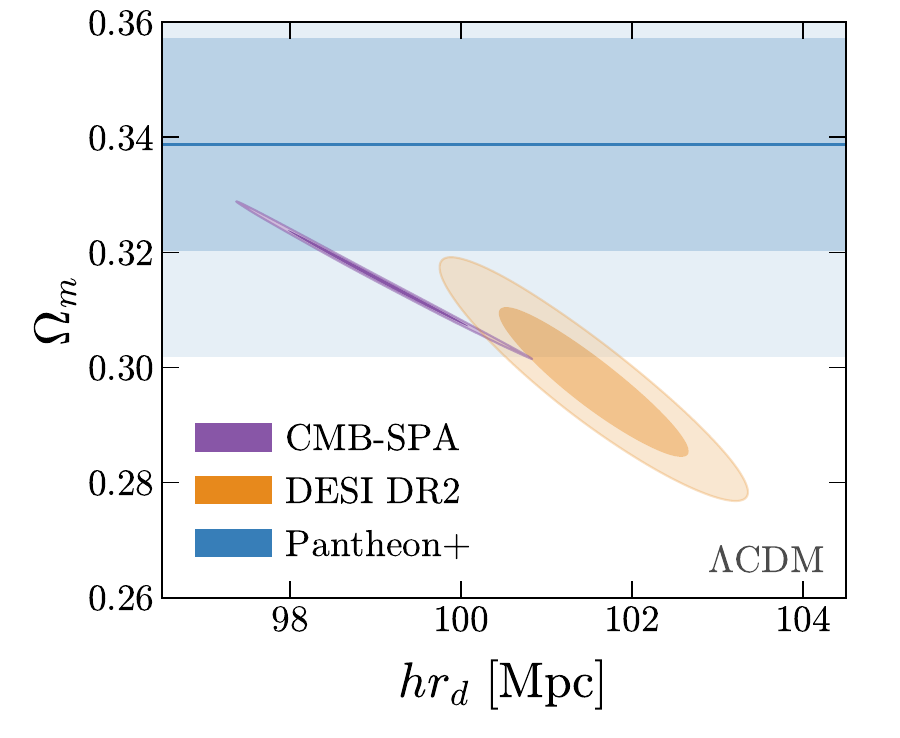}
    \caption{\emph{Left:} The mean and $1\sigma$ uncertainty of the posterior distributions of $\Omega_m$ in the flat $\Lambda$CDM model from the CMB, BAO, and SN~Ia data compilations. For Pantheon+ and DES-Dovekie, flux averaging increases the uncertainties, and shifts $\Omega_m$ towards a higher (lower) value for Pantheon+ (DES-Dovekie). Union3.1 is pre-binned at the catalog level and cannot be analyzed with flux averaging. \emph{Right:} Constraints in the $h r_d$--$\Omega_m$ plane from CMB-SPA, DESI~DR2 BAO, and Pantheon+ under flat $\Lambda$CDM. Inner (outer) contours show the 68\% (95\%) C.L. ranges. CMB-SPA and DESI~DR2 BAO differ at $2.5\sigma$ in the $h r_d$--$\Omega_m$ plane, and Pantheon+ deviates from CMB-SPA and DESI~DR2 BAO by $1.2\sigma$ and $2.0\sigma$ in $\Omega_m$.}
    \label{fig:Om_flatLCDM}
\end{figure*}

We first present the results of $\Omega_m$ from each probe separately under flat $\Lambda$CDM. The tensions observed between probes in this section motivate the extension of the dark energy model to absorb them. Figure~\ref{fig:Om_flatLCDM} and Table~\ref{tab:CPL} summarize the marginalized constraints. The $\Lambda$CDM constraints from CMB-SPA, DESI DR2 BAO, and SNe~Ia are obtained from our own MCMC chains, and are consistent with the official results \cite{Camphuis2026, DESIDR2, Hoyt2026, Dovekie2025}. We further apply flux averaging to Pantheon+ and DES-Dovekie to test the robustness of the results against weak-lensing magnification.

As shown in the left panel of \reffig{Om_flatLCDM}, CMB-SPA alone gives $\Omega_m = 0.3154 \pm 0.0055$, combining Planck with ACT~DR6 and SPT-3G. The three SN~Ia compilations prefer higher values, from $0.3292 \pm 0.0156$ (DES-Dovekie) to $0.3435 \pm 0.0250$ (Union3.1). DESI~DR2 BAO alone prefers a lower value, $\Omega_m = 0.2975 \pm 0.0086$.  SNe~Ia give weaker constraints on $\Omega_m$ than BAO or CMB. The three SN~Ia compilations are mutually consistent, and are also consistent with CMB within $1.2\sigma$. For Pantheon+ and DES-Dovekie, flux averaging increases the uncertainties, and shifts $\Omega_m$ towards a higher (lower) value for Pantheon+ (DES-Dovekie). The difference from DESI~DR2 BAO decreases from $2.0\sigma$ to $1.8\sigma$ for Pantheon+, and from $1.8\sigma$ to $1.0\sigma$ for DES-Dovekie
(with $\Omega_m$ also shifted toward the lower, BAO value). 

Because BAO measures distances in units of $r_d$, BAO alone can only constrain $\Omega_m$ and $hr_d$ in $\Lambda$CDM. The complete comparison between BAO and CMB is thus two-dimensional. The right panel of Figure~\ref{fig:Om_flatLCDM} shows the constraints in the $h r_d$--$\Omega_m$ plane, where $h r_d$ from CMB is derived from the CMB-SPA chains. The CMB constrains $\Omega_m$ and $h r_d$ with a strong anti-correlation, and DESI~DR2 BAO is displaced from the CMB along this degeneracy direction, toward lower $\Omega_m$ and higher $h r_d$, at $2.5\sigma$, which is larger than the $1.8\sigma$ difference in $\Omega_m$ alone. SNe~Ia constrain only $\Omega_m$, since $M_B$ and $H_0$ are analytically marginalized over. We show Pantheon+ in the right panel of Figure~\ref{fig:Om_flatLCDM} as an example. The tension between probes on this $h r_d$--$\Omega_m$ plane is the starting point for the discussion in Section~\ref{sec:disc_Om_tension}.

\subsection{Flat \texorpdfstring{$w_0 w_a$}{w0 wa}CDM}
\label{subsec:res_flatCPL}

\begin{figure*}[t]
    \centering
    \includegraphics[width=1.0\linewidth]{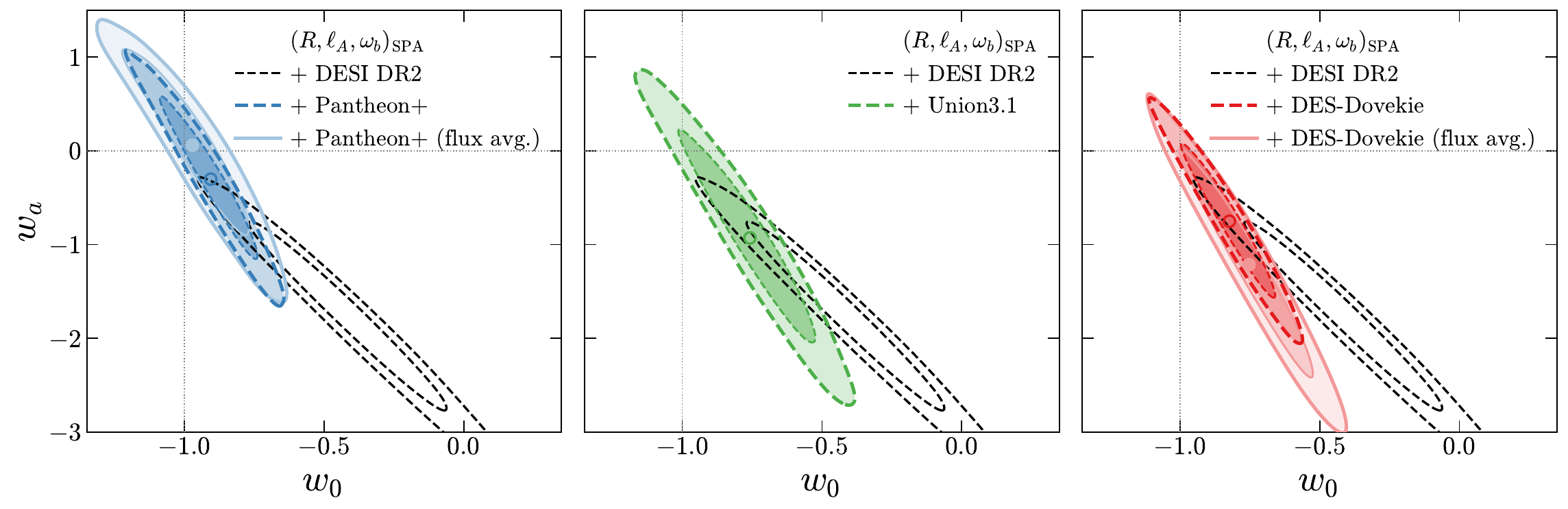}
    \includegraphics[width=1.0\linewidth]{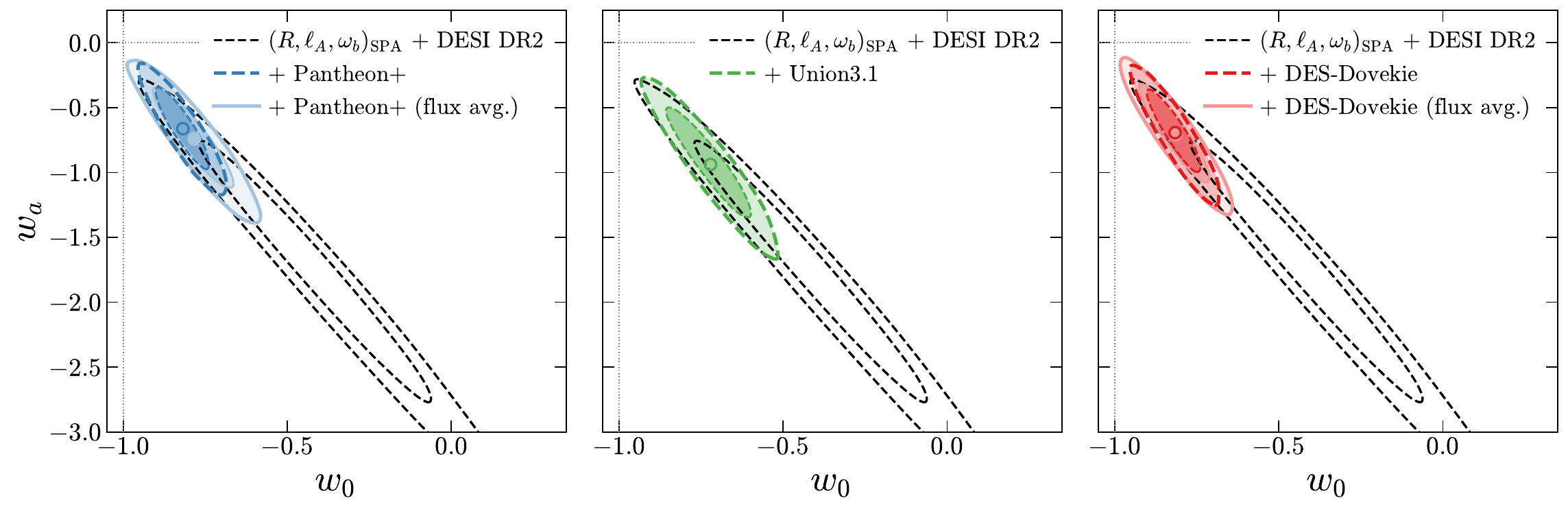}
    \caption{Constraints in the $w_0$--$w_a$ plane from different combinations of datasets. Each column corresponds to one SN~Ia compilation (Pantheon+, Union3.1, and DES-Dovekie, from left to right). Upper panels: the CMB-SPA distance prior $(R,\ell_A,\omega_b)_{\rm SPA}$ combined with each SN~Ia compilation (colored), overlaid with $(R,\ell_A,\omega_b)_{\rm SPA}$ + DESI DR2 (black dashed). Lower panels: $(R,\ell_A,\omega_b)_{\rm SPA}$ + DESI DR2 combined with each SN~Ia compilation (colored), with $(R,\ell_A,\omega_b)_{\rm SPA}$ + DESI DR2 shown in black dashed for reference. Light solid (dark dashed) colored contours show the flux-averaged (non-flux-averaged) results; Union3.1 is pre-binned at the catalog level, so only the non-flux-averaged result is shown. Inner and outer contours show the 68\% and 95\% C.L. ranges.}
    \label{fig:CPL_w0wa}
\end{figure*}
The most widely used extension of $\Lambda$CDM is the $w_0w_a$CDM model (see Equation~\eqref{eq:w0wa}). Since no single probe alone tightly constrains $(w_0, w_a)$ \cite{DESIDR2}, we report constraints from the combinations CMB + BAO, CMB + SNe Ia, and CMB + BAO + SNe~Ia, using the CMB-SPA distance prior $(R,\ell_A,\omega_b)_{\rm SPA}$ presented in Section~\ref{subsec:data_cmb}. The numerical constraints are summarized in Table~\ref{tab:CPL}.

To quantify the preference for a departure from $\Lambda$CDM, we obtain the best-fit $\chi^2$ of both models by numerical minimization and compute $\Delta\chi^2_{\rm MAP} \equiv \chi^2_{\Lambda{\rm CDM}} - \chi^2_{w_0w_a{\rm CDM}}$, where ``MAP'' stands for ``maximum a posteriori''. Under Wilks' theorem, $\Delta\chi^2_{\rm MAP}$ follows a $\chi^2$ distribution with two degrees of freedom when $\Lambda$CDM is true, and we convert the corresponding $p$-value into a two-sided Gaussian-equivalent significance. This is the same diagnostic tool used in DESI DR2 BAO analysis \cite{DESIDR2}. The values of $\Delta\chi^2_{\rm MAP}$ and the corresponding significances are listed in Table~\ref{tab:dchi2}.

\begin{table*}[t]
\centering
\caption{Summary of cosmological parameter constraints from the CMB, BAO, and SN~Ia datasets in flat $\Lambda$CDM and flat $w_0w_a$CDM. We quote the marginalized mean and the 68\% C.L. ranges. In flat $w_0w_a$CDM, the CMB information enters through the distance prior $(R,\ell_A,\omega_b)_{\rm SPA}$. SN~Ia constraints are quoted with and without flux averaging.}
\label{tab:CPL}
\small
\renewcommand{\arraystretch}{1.3}
\begin{tabular*}{\textwidth}{@{\extracolsep{\fill}}lccc}
\toprule
Model / Dataset & $\Omega_m$ & $w_0$ & $w_a$ \\
\hline
\midrule
\multicolumn{4}{l}{\textbf{flat $\Lambda$CDM}} \\
\midrule
CMB-SPA               & $0.3154 \pm 0.0055$       & --- & --- \\
DESI DR2              & $0.2975 \pm 0.0086$       & --- & --- \\
\addlinespace[3pt]
Pantheon+           & $0.3388 \pm 0.0185$ & --- & --- \\
Pantheon+ (flux avg.) & $0.3551 \pm 0.0303$ & --- & --- \\
\addlinespace[3pt]
Union3.1                & $0.3435 \pm 0.0250$ & --- & --- \\
\addlinespace[3pt]
DES-Dovekie           & $0.3292 \pm 0.0156$ & --- & --- \\
DES-Dovekie (flux avg.
) & $0.3191 \pm 0.0211$ & --- & --- \\
\midrule
\hline
\multicolumn{4}{l}{\textbf{flat $w_0 w_a$CDM}} \\
\midrule
$(R,\ell_A,\omega_b)_{\rm SPA}$ $+$ Pantheon+             & $0.3218 \pm 0.0134$ & $-0.906 \pm 0.114$ & $-0.302^{+0.563}_{-0.565}$ \\
$(R,\ell_A,\omega_b)_{\rm SPA}$ $+$ Pantheon+ (flux avg.)  & $0.3338^{+0.0216}_{-0.0211}$ & $-0.972^{+0.147}_{-0.149}$ & $+0.059^{+0.686}_{-0.670}$ \\
\addlinespace[3pt]
$(R,\ell_A,\omega_b)_{\rm SPA}$ $+$ DES-Dovekie            & $0.3127^{+0.0095}_{-0.0093}$ & $-0.824 \pm 0.111$ & $-0.751^{+0.538}_{-0.540}$ \\
$(R,\ell_A,\omega_b)_{\rm SPA}$ $+$ DES-Dovekie (flux avg.) & $0.3037^{+0.0121}_{-0.0118}$ & $-0.753 \pm 0.155$ & $-1.215^{+0.799}_{-0.801}$ \\
\addlinespace[3pt]
$(R,\ell_A,\omega_b)_{\rm SPA}$ $+$ Union3.1                  & $0.3194^{+0.0132}_{-0.0137}$ & $-0.758^{+0.159}_{-0.157}$ & $-0.929 \pm 0.738$ \\
\midrule
$(R,\ell_A,\omega_b)_{\rm SPA}$ $+$ DESI DR2                & $0.3550^{+0.0240}_{-0.0236}$ & $-0.394^{+0.238}_{-0.240}$ & $-1.818^{+0.685}_{-0.675}$ \\
\midrule
\addlinespace[3pt]
$(R,\ell_A,\omega_b)_{\rm SPA}$ $+$ DESI DR2 + Pantheon+                & $0.3137^{+0.0054}_{-0.0055}$ & $-0.818^{+0.055}_{-0.053}$ & $-0.663^{+0.203}_{-0.207}$ \\
$(R,\ell_A,\omega_b)_{\rm SPA}$ $+$ DESI DR2 + Pantheon+ (flux avg.)   & $0.3175^{+0.0084}_{-0.0082}$ & $-0.784^{+0.080}_{-0.081}$ & $-0.742^{+0.261}_{-0.254}$ \\
\addlinespace[3pt]
$(R,\ell_A,\omega_b)_{\rm SPA}$ $+$ DESI DR2 + DES-Dovekie              & $0.3129 \pm 0.0052$ & $-0.816^{+0.052}_{-0.053}$ & $-0.693^{+0.208}_{-0.203}$ \\
$(R,\ell_A,\omega_b)_{\rm SPA}$ $+$ DESI DR2 + DES-Dovekie (flux avg.
)& $0.3124 \pm 0.0068$ & $-0.815^{+0.070}_{-0.071}$ & $-0.708^{+0.249}_{-0.250}$ \\
\addlinespace[3pt]
$(R,\ell_A,\omega_b)_{\rm SPA}$ $+$ DESI DR2 + Union3.1                   & $0.3223 \pm 0.0081$ & $-0.721^{+0.083}_{-0.085}$ & $-0.938^{+0.279}_{-0.275}$ \\
\bottomrule
\end{tabular*}
\end{table*}

The upper panels of Figure~\ref{fig:CPL_w0wa} show the marginalized constraints in the $w_0$--$w_a$ plane from CMB+BAO, and from CMB combined with each of the three SN~Ia compilations, with and without flux averaging. The CMB+BAO combination gives a $2.90\sigma$ preference for a departure from $\Lambda$CDM, while the CMB+SN~Ia combinations all deviate by at most $1.16\sigma$ ($0.95\sigma$, $1.16\sigma$, and $1.08\sigma$ for Pantheon+, Union3.1, and DES-Dovekie). Flux averaging further reduces the preference for deviation from $\Lambda$CDM for Pantheon+ and DES-Dovekie by increasing their uncertainties.

The lower panels of Figure~\ref{fig:CPL_w0wa} present the combination of CMB, BAO, and each of the three SN~Ia compilations. As the upper panels show, the SN~Ia constraints have a degeneracy direction different from that of CMB+BAO, so the combination of three probes further shrinks the contours. For Pantheon+, Union3.1, and DES-Dovekie, the preference for a departure from $\Lambda$CDM reaches $2.98\sigma$, $3.18\sigma$, and $3.05\sigma$, respectively. With flux averaging, the preference for a departure from $\Lambda$CDM for the Pantheon+ and DES-Dovekie combinations reduces to $2.53\sigma$ and $2.61\sigma$. The centers of the $(w_0, w_a)$ posteriors move mainly along the degeneracy direction, by $0.3\sigma$ (Pantheon+) and $0.01\sigma$ (DES-Dovekie) in the two-dimensional plane, so the decrease of the preference comes mainly from the larger uncertainties.

These results follow the pattern of the $\Lambda$CDM tensions in Section~\ref{subsec:res_flatLCDM}. The preference for a departure from $\Lambda$CDM from CMB+BAO is $2.90\sigma$. This reflects the tension between BAO and the CMB in the $h r_d$--$\Omega_m$ plane (right panel of Figure~\ref{fig:Om_flatLCDM}). The SN~Ia compilations differ from the CMB only in $\Omega_m$ by at most $1.2\sigma$, and CMB+SN~Ia correspondingly gives at most $1.16\sigma$ significance of deviation from $\Lambda$CDM model. Once BAO is included, the results are highly consistent among the three SN~Ia compilations. They select nearly the same segment along the CMB+BAO degeneracy direction in the $w_0$--$w_a$ plane (lower panels of Figure~\ref{fig:CPL_w0wa}), and give $2.98\sigma$, $3.18\sigma$, and $3.05\sigma$ significance for a deviation from $\Lambda$CDM for Pantheon+, Union3.1, and DES-Dovekie, respectively. With the updated SN~Ia compilations, the choice of SN~Ia sample therefore changes the preference for evolving dark energy by only $0.2\sigma$, which is much smaller than the spread from $2.8\sigma$ to $4.2\sigma$ reported with the original compilations \cite{DESIDR2}. Our result is consistent with Ref.~\cite{Hoyt2026}, which also found that the difference among the updated SN~Ia compilations in the CMB+BAO+SN~Ia analysis is around 0.2$\sigma$.

\subsection{Model-independent dark energy density evolution}
\label{subsec:res_Xz}
We present the results of model-independent constraints on dark energy density ratio $X(z) = \rho_{\rm DE}(z)/\rho_{\rm DE}(0)$ at five redshifts by combining the CMB, BAO, and the SN~Ia compilations with and without flux averaging. The numerical results are summarized in Table~\ref{tab:Xz}. 

\begin{table*}[!t]
\centering
\caption{Reconstructed $X(z)$ knot values from the CMB-SPA distance prior $(R,\ell_A,\omega_b)_{\rm SPA}$ combined with DESI DR2 BAO, with and without SNe~Ia. Values are posterior means with 68\% C.L. The bottom block replaces the distance prior with the full CMB-SPA power spectra likelihood, with and without CMB lensing, for DESI DR2 + Pantheon+.}
\label{tab:Xz}
\newcommand{\xv}[2]{$#1 \pm #2$}
\newcommand{\xvasym}[3]{$#1^{+#2}_{-#3}$}
\small
\renewcommand{\arraystretch}{1.4}
\setlength{\tabcolsep}{4pt}
\resizebox{\textwidth}{!}{%
\begin{tabular}{@{}lccccccc@{}}
\toprule
Dataset & $\Omega_m$ & \makecell{$H_0$\\[-1pt]{\footnotesize [km\,s$^{-1}$\,Mpc$^{-1}$]}} & $X(1/3)$ & $X(2/3)$ & $X(1)$ & $X(4/3)$ & $X(2.33)$ \\
\midrule
$(R,\ell_A,\omega_b)_{\rm SPA}$ + DESI DR2 & \xvasym{0.272}{0.099}{0.096} & \xvasym{76.1}{13.8}{14.1} & \xvasym{0.96}{0.55}{0.61} & \xvasym{1.09}{0.43}{0.49} & \xvasym{0.84}{0.41}{0.45} & \xvasym{0.74}{0.34}{0.38} & \xvasym{0.56}{0.32}{0.35} \\
\addlinespace[2pt]
\quad $+$ Pantheon+ & \xv{0.313}{0.006} & \xv{67.5}{0.6} & \xv{1.08}{0.03} & \xv{1.15}{0.06} & \xv{0.96}{0.10} & \xv{0.82}{0.16} & \xv{0.63}{0.27} \\
\quad $+$ Pantheon+ (flux avg.) & \xv{0.315}{0.008} & \xv{67.3}{0.9} & \xv{1.09}{0.05} & \xv{1.20}{0.08} & \xv{0.96}{0.11} & \xv{0.83}{0.16} & \xv{0.64}{0.27} \\
\addlinespace[2pt]
\quad $+$ Union3.1 & \xv{0.319}{0.009} & \xv{66.8}{0.9} & \xv{1.12}{0.06} & \xvasym{1.16}{0.07}{0.06} & \xv{1.03}{0.11} & \xv{0.85}{0.16} & \xvasym{0.66}{0.27}{0.28} \\
\addlinespace[2pt]
\quad $+$ DES-Dovekie & \xv{0.313}{0.006} & \xv{67.5}{0.6} & \xv{1.07}{0.03} & \xv{1.08}{0.05} & \xv{1.04}{0.10} & \xv{0.81}{0.15} & \xv{0.64}{0.27} \\
\quad $+$ DES-Dovekie (flux avg.) & \xvasym{0.312}{0.007}{0.008} & \xv{67.6}{0.8} & \xvasym{1.07}{0.04}{0.05} & \xv{1.08}{0.06} & \xv{1.02}{0.11} & \xv{0.81}{0.15} & \xv{0.64}{0.27} \\
\midrule
\multicolumn{8}{@{}l}{DESI DR2 + Pantheon+} \\
\quad $+$ CMB-SPA & \xv{0.313}{0.006} & \xv{67.5}{0.6} & \xv{1.08}{0.03} & \xv{1.15}{0.06} & \xv{0.96}{0.10} & \xv{0.83}{0.15} & \xvasym{0.61}{0.24}{0.25} \\
\quad $+$ CMB-SPA (no lensing) & \xv{0.313}{0.006} & \xv{67.5}{0.6} & \xv{1.09}{0.03} & \xv{1.15}{0.06} & \xv{0.96}{0.10} & \xv{0.84}{0.16} & \xvasym{0.62}{0.26}{0.27} \\
\bottomrule
\end{tabular}}

\vspace{1.5em}

\captionsetup{justification=Justified}   
\caption{Preference for a departure from $\Lambda$CDM in flat $w_0w_a$CDM and flat $X(z)$CDM. $\Delta\chi^2_{\rm MAP} \equiv \chi^2_{\Lambda{\rm CDM}} - \chi^2_{\rm model}$ is evaluated at the maximum a posteriori (MAP) points obtained by numerical minimization with identical likelihoods. The significance is the two-sided Gaussian-equivalent of the $p$-value under Wilks' theorem, with two ($w_0w_a$CDM) and five ($X(z)$CDM) additional degrees of freedom. The values in parentheses decompose $\Delta\chi^2_{\rm MAP}$ into the contributions of the CMB, BAO, and SN~Ia likelihoods, $\Delta\chi^2_i \equiv \chi^2_{i,\Lambda{\rm CDM}} - \chi^2_{i,{\rm model}}$, evaluated at the respective best fits.}
\label{tab:dchi2}
\centering
\small
\renewcommand{\arraystretch}{1.2}
\setlength{\tabcolsep}{4pt}
\scalebox{1}{%
\begin{tabular}{@{}lcccccc@{}}
\toprule
 & \multicolumn{3}{c}{flat $w_0w_a$CDM} & \multicolumn{3}{c}{flat $X(z)$CDM} \\
\cmidrule(lr){2-4}\cmidrule(l){5-7}
Dataset & $\Delta\chi^2_{\rm MAP}$ & Significance & \makecell{$(\Delta\chi^2_{\rm CMB},\,\Delta\chi^2_{\rm BAO},$\\$\Delta\chi^2_{\rm SN})$} & $\Delta\chi^2_{\rm MAP}$ & Significance & \makecell{$(\Delta\chi^2_{\rm CMB},\,\Delta\chi^2_{\rm BAO},$\\$\Delta\chi^2_{\rm SN})$} \\
\midrule
$(R,\ell_A,\omega_b)_{\rm SPA}$ $+$ DESI DR2 & 11.15 & $2.90\sigma$ & (4.91, 6.25, ---) & 12.51 & $2.19\sigma$ & (4.66, 7.85, ---) \\
\quad $+$ Pantheon+ & 11.67 & $2.98\sigma$ & (3.22, 4.74, 3.72) & 16.32 & $2.75\sigma$ & (3.69, 8.04, 4.59) \\
\quad $+$ Pantheon+ (flux avg.) & 8.95 & $2.53\sigma$ & (3.76, 4.65, 0.55) & 15.07 & $2.57\sigma$ & (4.19, 7.95, 2.93) \\
\quad $+$ Union3.1 & 13.01 & $3.18\sigma$ & (3.99, 5.29, 3.72) & 13.71 & $2.38\sigma$ & (4.12, 6.73, 2.87) \\
\quad $+$ DES-Dovekie & 12.17 & $3.05\sigma$ & (3.44, 4.49, 4.24) & 12.88 & $2.25\sigma$ & (3.72, 4.99, 4.17) \\
\quad $+$ DES-Dovekie (flux avg.) & 9.41 & $2.61\sigma$ & (4.18, 3.57, 1.67) & 10.00 & $1.78\sigma$ & (4.40, 4.69, 0.91) \\
\midrule
\multicolumn{7}{@{}l}{$(R,\ell_A,\omega_b)_{\rm SPA}$} \\
\quad $+$ Pantheon+ & 2.14 & $0.95\sigma$ & (0.15, ---, 1.99) & --- & --- & --- \\
\quad $+$ Pantheon+ (flux avg.) & 1.46 & $0.70\sigma$ & (0.05, ---, 1.41) & --- & --- & --- \\
\quad $+$ Union3.1 & 2.79 & $1.16\sigma$ & (0.05, ---, 2.74) & --- & --- & --- \\
\quad $+$ DES-Dovekie & 2.54 & $1.08\sigma$ & (0.08, ---, 2.46) & --- & --- & --- \\
\quad $+$ DES-Dovekie (flux avg.) & 2.04 & $0.91\sigma$ & (0.00, ---, 2.04) & --- & --- & --- \\
\bottomrule
\end{tabular}}
\end{table*}

\begin{figure*}[tp]
    \centering
    \includegraphics[width=1.0\linewidth]{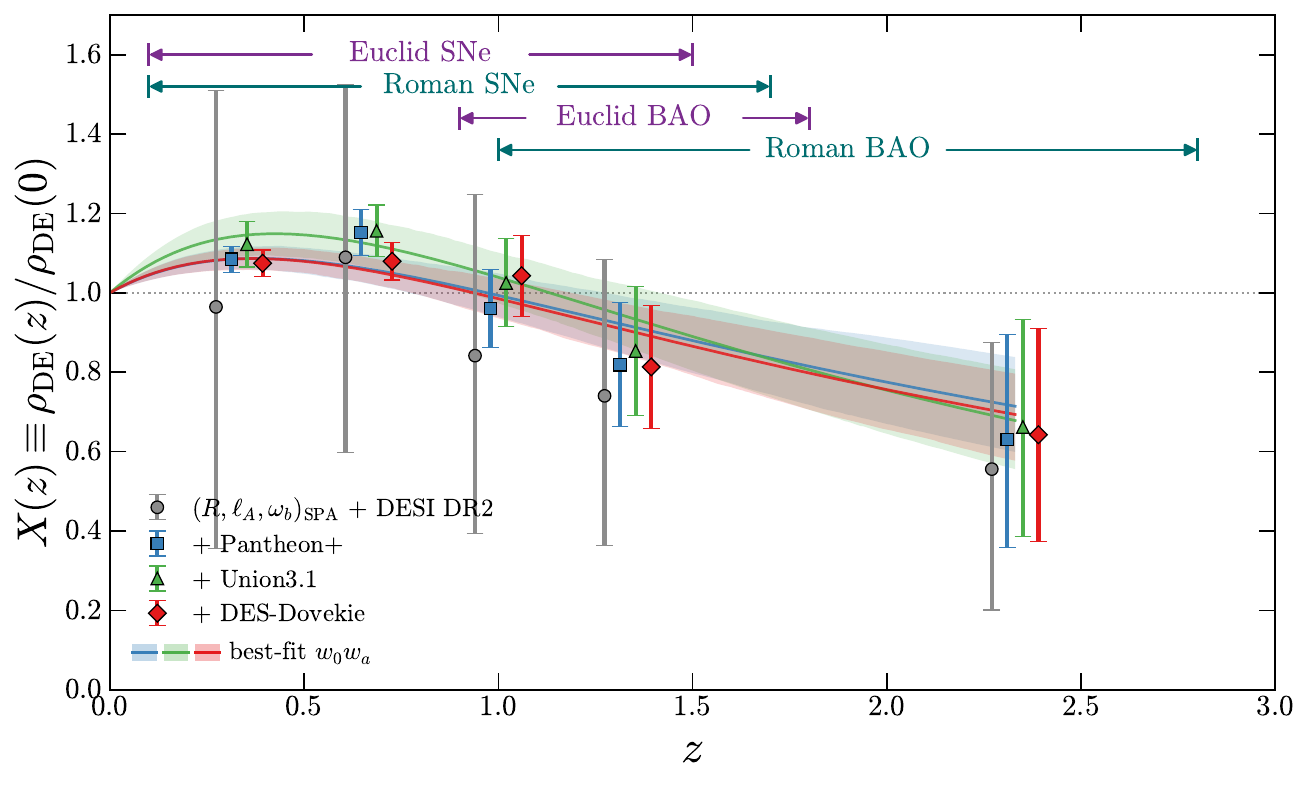}
    \caption{Model-independent constraints on $X(z) = \rho_{\rm DE}(z)/\rho_{\rm DE}(0)$ from $(R,\ell_A,\omega_b)_{\rm SPA}$ + DESI DR2, with and without SN~Ia (not flux-averaged). Error bars show the 68\% C.L. ranges at the five knots. The knots are offset horizontally for visualization. The shaded bands are the 68\% C.L. ranges from the corresponding $w_0w_a$CDM posterior. The gray points show $(R,\ell_A,\omega_b)_{\rm SPA}$ + DESI DR2 alone, whose posteriors are converted from the sampled {$\omega_{\rm DE}(z_k) = \Omega_{\rm DE} h^2 X(z_k)$}. The dotted line marks the cosmological constant, $X = 1$. The redshift coverage of the Euclid \cite{Astier2014, Laureijs2011} and Roman \cite{Hounsell2018, Wang2022HLSS} SN~Ia and BAO surveys is indicated at the top.}
    \label{fig:Xz_main}
    \vspace{12pt}
    \includegraphics[width=1.0\linewidth]{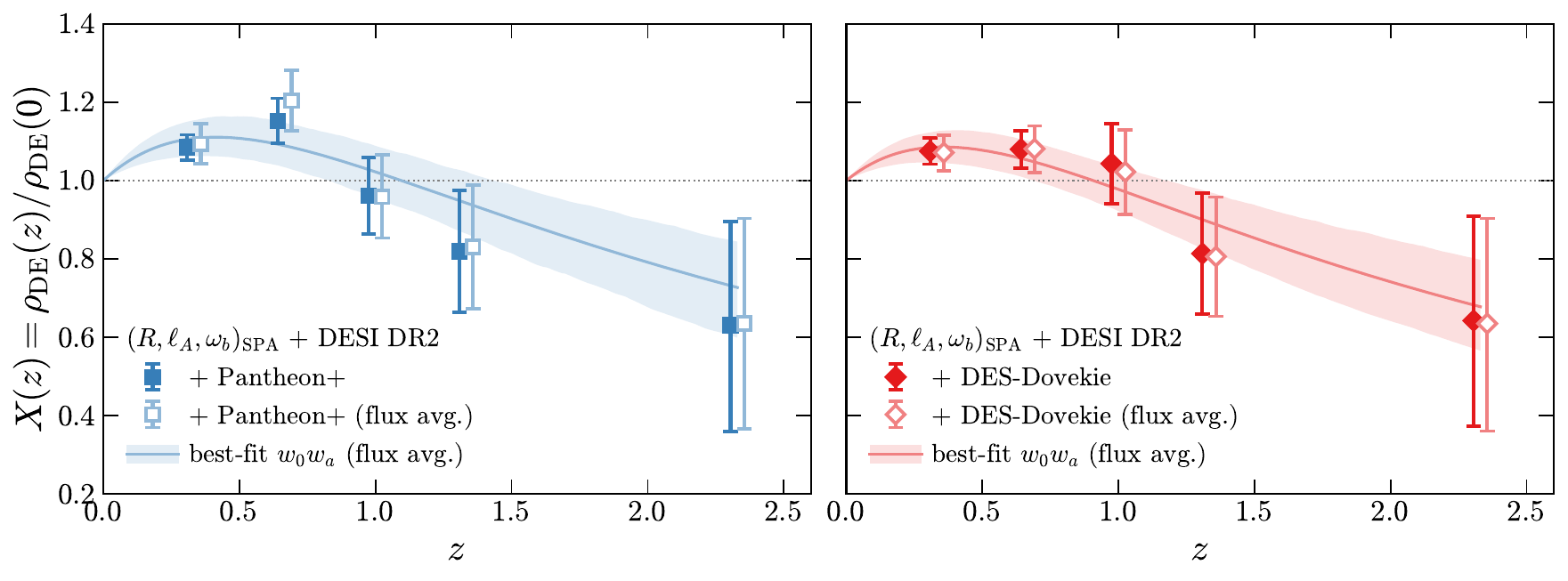}
    \caption{Effect of flux averaging the SNe~Ia on $X(z)$ from $(R,\ell_A,\omega_b)_{\rm SPA}$ + DESI DR2 combined with Pantheon+ (left) and DES-Dovekie (right). Filled (open) markers are without (with) flux averaging, and the error bars show the {68\% C.L. ranges} at the five knots. Shaded bands show the 68\% C.L. ranges from the flux-averaged $w_0w_a$CDM posterior, and the dotted line marks $X = 1$.}
    \label{fig:Xz_FA}
\end{figure*}

Figure~\ref{fig:Xz_main} shows the $X(z)$ obtained from CMB-SPA distance prior, DESI DR2 BAO, combined with each of the three SN~Ia compilations without flux averaging. As a consistency check, we plot the marginalized mean and 68\% C.L. ranges of $X(z)$ converted from the corresponding $w_0w_a$CDM chains using the same datasets as the solid lines and colored shades, where Pantheon+ and DES-Dovekie nearly overlap at $z < 1$. For each combination, the $X(z)$ from the model-independent approach generally agrees well with the $w_0w_a$ model. The dark energy density first increases from $z = 0$ and then starts to decrease around $z\sim 2/3$, and keeps dropping after crossing $X(z)=1$ around $z = 1$. The uncertainties of the model-independent $X(z)$ knots are larger than those of the density converted from $w_0w_a$. This difference in uncertainties is expected, since $X(z)$ at five knots are allowed to evolve freely, which means we have more degrees of freedom to constrain from the same datasets. For all chains obtained from CMB+BAO+SN~Ia, the uncertainties of $X(z)$ grow with redshift. This is simply because the number of SNe~Ia drops rapidly at higher redshifts and the constraining power on $X(z)$ at high redshift mainly comes from the CMB and BAO.

\begin{figure*}[t]
    \centering
    \includegraphics[width=0.85\linewidth]{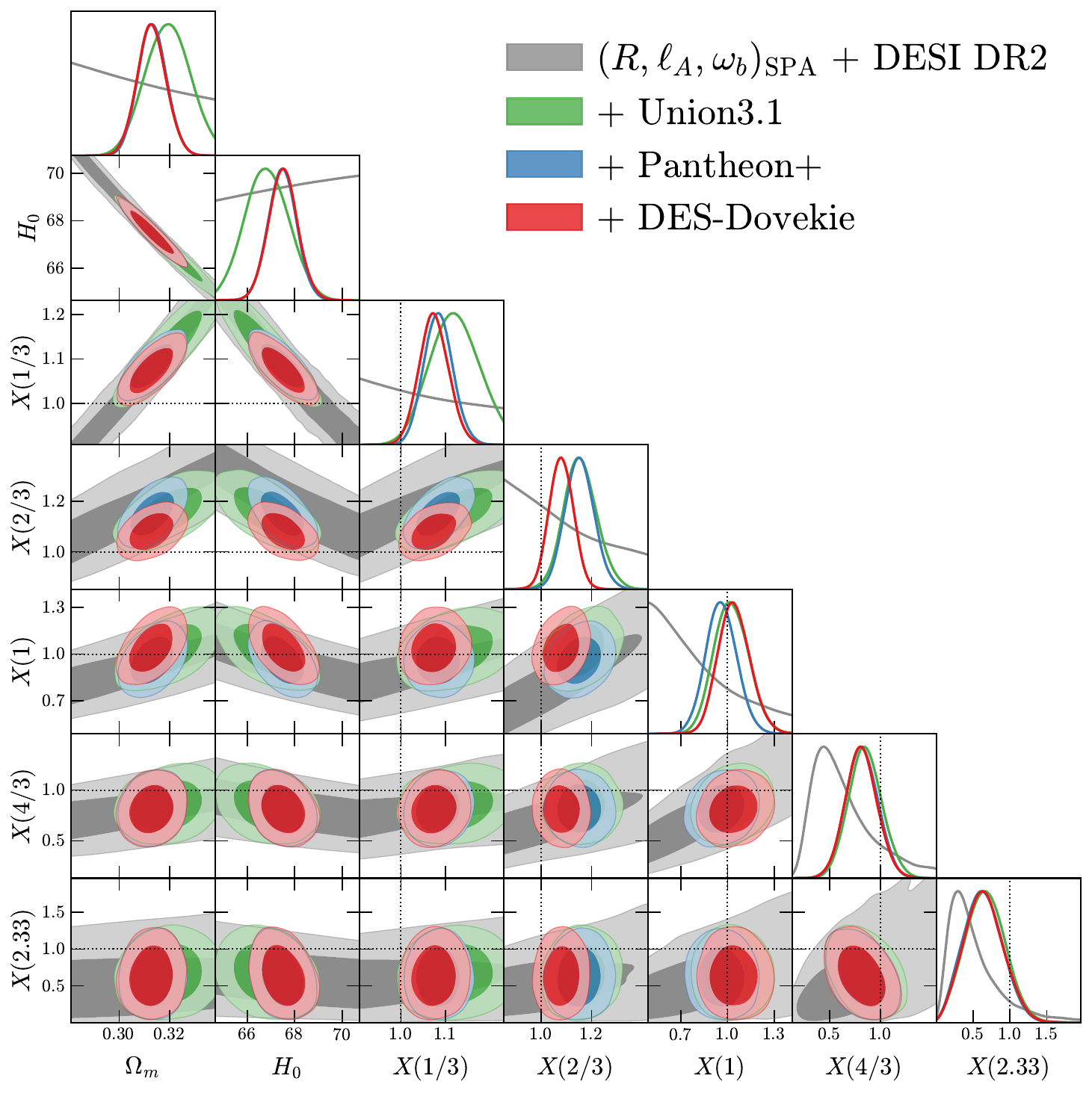}
    \caption{Contours of the model-independent constraints on $X(z) = \rho_{\rm DE}(z)/\rho_{\rm DE}(0)$ from CMB, BAO and SNe~Ia without flux averaging. The gray contours and lines represent the result for $(R,\ell_A,\omega_b)_{\rm SPA}$ + DESI DR2. The colored contours represent the result after including the Pantheon+ (blue), Union3.1 (green), and DES-Dovekie (red). Inner (outer) contours {show the 68\% (95\%) C.L. ranges.} The dark dotted lines mark the cosmological constant $X(z) = 1$.
    }
    \label{fig:Xz_contour_SNe}
\end{figure*}

\begin{figure}[t]
    \centering
    \includegraphics[width=\linewidth]{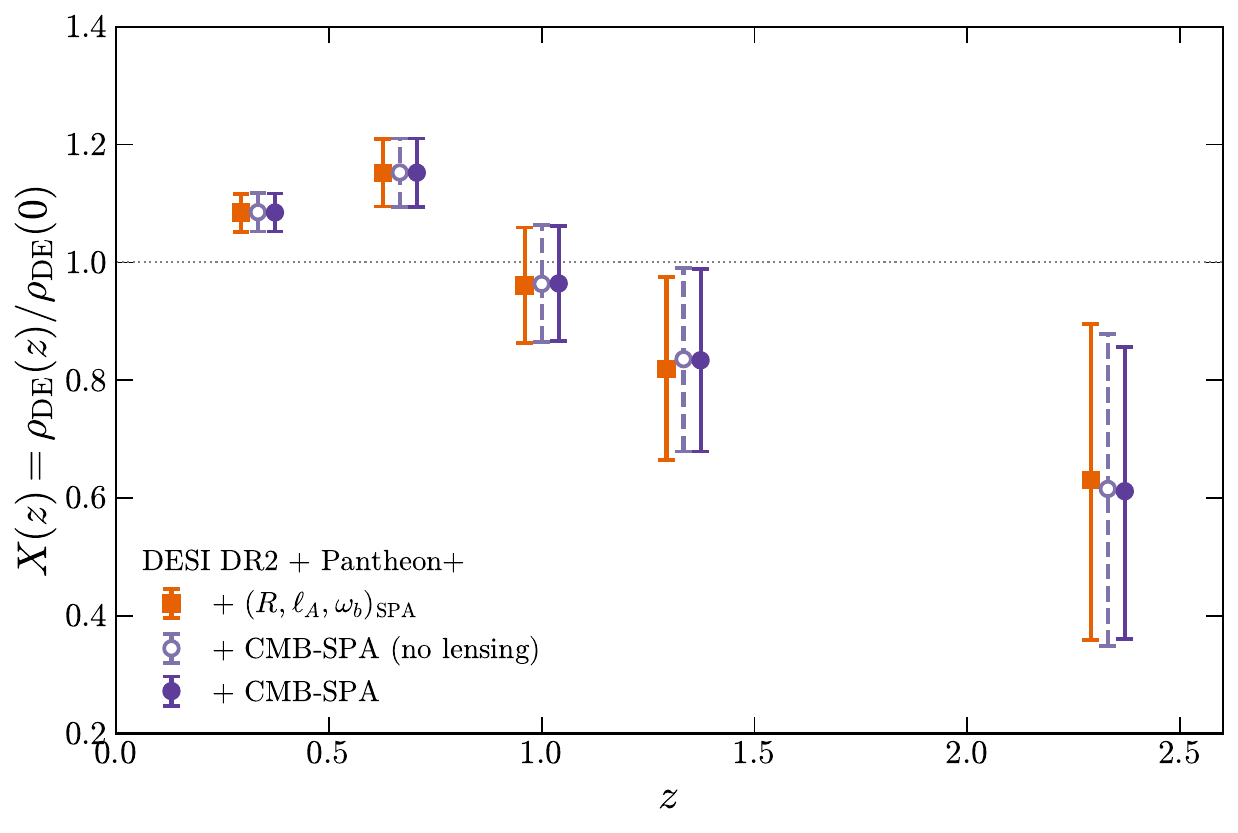}
    \caption{Dependence of the constrained $X(z)$ on the treatment of the CMB, for DESI~DR2 BAO combined with Pantheon+ without flux averaging. Orange squares use the CMB-SPA distance prior $(R, \ell_A, \omega_b)$. Filled purple and open light purple circles (dashed error bars) use the full CMB-SPA temperature and polarization power spectra with and without the CMB lensing reconstruction likelihood. Points at the same knot are offset horizontally for visualization. The dotted line represents $X(z) = 1$. The three treatments agree to within $0.11\sigma$ at all knots.}
    \label{fig:Xz_CMB_comparison}
\end{figure}

\begin{figure}[h]
    \centering
    \includegraphics[width=\linewidth]{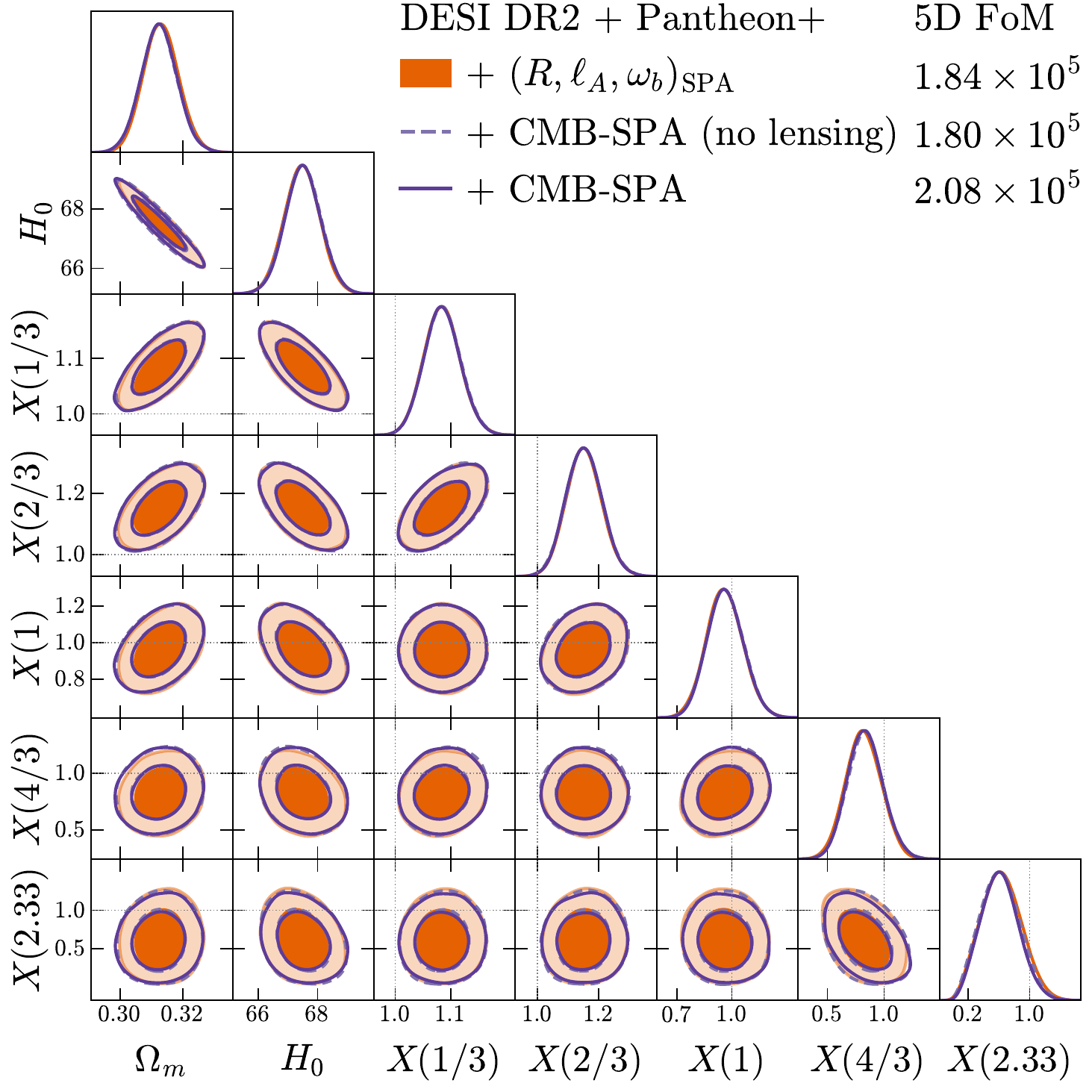}
    \caption{Comparison of our constraints on $X(z)$CDM from DESI DR2 BAO, Pantheon+ (without flux averaging) and different CMB likelihoods. The filled orange contours use the CMB-SPA distance prior $(R,\ell_A,\omega_b)_{\rm SPA}$. The purple solid (light purple dashed) lines use the full CMB-SPA power spectra with (without) the CMB lensing reconstruction likelihood. Inner (outer) contours show the 68\% (95\%) C.L. ranges. The dark dotted lines mark the cosmological constant $X(z) = 1$. The column to the right of the legend gives the figure of merit of the five $X(z)$ knots, ${\rm FoM} = [\det \mathbf{C}(X)]^{-1/2}$ \cite{Albrecht2006, Wang2008FoM}.}
    \label{fig:Xz_CMB_comparison_contour}
\end{figure}

Without SN~Ia, the $X(z)$ posteriors from CMB+BAO are highly skewed toward large $X$, as shown in \reffig{Xz_contour_SNe}, because CMB and BAO constrain $\omega_{\rm DE}(z_k)$ but only weakly constrain $\omega_{\rm DE}(0)$ (Section~\ref{sec:Xz_method}). The posterior means quoted in Table~\ref{tab:Xz} are pulled toward this tail, whereas the medians, $X = 0.70,\,0.89,\,0.66,\,0.59,\,0.42$, are close to the best-fit values and lie below unity at all knots. SNe~Ia constrain $\Omega_m(1+z)^3 + (1-\Omega_m)X(z)$ at many redshifts, which breaks the degeneracy between $\Omega_m$, $X(z)$, and $h$, and thus tightly constrains each of them.

Table~\ref{tab:dchi2} quantifies the preference for a departure from $\Lambda$CDM in the $X(z)$ constraints and compares it with $w_0w_a$CDM. Since $X(z)$ has five instead of two additional degrees of freedom, the same improvement of the fit corresponds to a lower significance. For CMB+BAO, and for CMB+BAO combined with Union3.1 or DES-Dovekie, $X(z)$ improves the fit over $w_0w_a$CDM only marginally, by $\Delta\chi^2_{\rm MAP} = 1.36$, $0.70$, and $0.71$, respectively. The reconstruction therefore captures essentially the same departure from $\Lambda$CDM as $w_0w_a$CDM, and the significance decreases accordingly, from $2.90\sigma$ to $2.19\sigma$ for CMB+BAO, from $3.18\sigma$ to $2.38\sigma$ for adding Union3.1, and from $3.05\sigma$ to $2.25\sigma$ for adding DES-Dovekie.

Pantheon+ behaves somewhat differently from the other two compilations. With Pantheon+, $X(z)$ improves the fit over $w_0w_a$CDM by $\Delta\chi^2_{\rm MAP} = 4.65$, so that Pantheon+ changes from the compilation with the lowest preference in $w_0w_a$CDM ($2.98\sigma$) to the one with the highest in $X(z)$ ($2.75\sigma$). Decomposing $\chi^2$ by probe at the best fits, most of this additional improvement comes from BAO ($\Delta\chi^2_{\rm BAO} = 3.30$), with smaller contributions from the SNe~Ia ($0.87$) and the CMB ($0.47$), whereas for Union3.1 and DES-Dovekie the BAO term improves by only $1.44$ and $0.50$ and the SN~Ia term does not improve. With three more free parameters than $w_0w_a$CDM, an improvement of $4.65$ is comparable to what is expected from the additional freedom alone and is therefore not statistically significant. It nevertheless indicates that the redshift dependence of the Pantheon+ distances differs from that of the other two compilations in a way that the two-parameter $w_0w_a$ form cannot model. The additional freedom is used mainly at $z = 2/3$, where the Pantheon+ reconstruction, $X(2/3) = 1.15 \pm 0.06$, lies above the $w_0w_a$CDM prediction by about one standard deviation, whereas the Union3.1 and DES-Dovekie knots agree with their $w_0w_a$CDM predictions (Figure~\ref{fig:Xz_main}). This local excess coincides with the redshift of the DESI~DR2 LRG2 measurement at $z_{\rm eff} = 0.706$, and is consistent with the BAO term dominating the improvement in $\chi^2$. Since the $X(z)$ knots and the $w_0w_a$CDM predictions are derived from the same data, this comparison localizes the redshift at which the additional improvement in the fit arises. 

Flux averaging sharpens the local feature of the results using Pantheon+ at $z \simeq 2/3$, but leaves the shape of $X(z)$ essentially unchanged for the results using DES-Dovekie (Figure~\ref{fig:Xz_FA}). For Pantheon+, it raises $X(2/3)$ from $1.15 \pm 0.06$ to $1.20 \pm 0.08$, while the knots at $z \geq 1$ are unchanged, so that the flux-averaged reconstruction peaks at $z \simeq 2/3$ and drops to $X(1) = 0.96$. The flux-averaged $w_0w_a$CDM fit, by contrast, gives $X(2/3) = 1.09^{+0.06}_{-0.05}$. In $w_0w_a$CDM, flux averaging moves $(w_0, w_a)$ by only $0.3\sigma$, but the SN~Ia contribution to the improvement over $\Lambda$CDM drops from $\Delta\chi^2_{\rm SN} = 3.72$ to $0.55$, and the significance decreases from $2.98\sigma$ to $2.53\sigma$. In $X(z)$, the SN~Ia term drops less, from $4.59$ to $2.93$, and the significance decreases only from $2.75\sigma$ to $2.57\sigma$, so that the improvement of $X(z)$ over $w_0w_a$CDM grows from $\Delta\chi^2_{\rm MAP} = 4.65$ to $6.12$. Flux averaging thus appears to remove the part of the Pantheon+ preference that a smooth $w_0w_a$ evolution can absorb, while retaining the local excess at $z \simeq 2/3$.

In contrast to Pantheon+, the shape of $X(z)$ for DES-Dovekie is not sensitive to flux averaging. The knots shift by at most $0.2\sigma$, with $X(2/3)$ staying at $1.08$ whereas it rises from $1.15$ to $1.20$ for Pantheon+. In Table~\ref{tab:dchi2}, $\Delta\chi^2_{\rm MAP}$ of $X(z)$ exceeds that of $w_0w_a$CDM by only $0.71$ without and $0.59$ with flux averaging. For both compilations, flux averaging inflates the uncertainties at $z \lesssim 2/3$ by factors of $1.2$--$1.6$. With these larger uncertainties, the SN~Ia contribution to the improvement of $X(z)$ over $\Lambda$CDM for DES-Dovekie drops from $\Delta\chi^2_{\rm SN} = 4.17$ to $0.91$, which lowers $\Delta\chi^2_{\rm MAP}$ from $12.88$ to $10.00$ and the significance from $2.25\sigma$ to $1.78\sigma$.

Figures~\ref{fig:Xz_CMB_comparison} and \ref{fig:Xz_CMB_comparison_contour} compare the CMB distance prior $(R,\ell_A,\omega_b)_{\rm SPA}$ with the full CMB-SPA power spectra, with and without the CMB lensing likelihood, for the joint analysis of $X(z)$CDM with DESI DR2 and Pantheon+ without flux averaging. The three treatments agree to within $0.11\sigma$ on $\Omega_m$, $H_0$, and all $X(z)$ knots, and the uncertainties at $z \leq 4/3$ agree to within $2\%$. At $z = 2.33$, the CMB lensing likelihood reduces the uncertainty of $X(2.33)$ by $6\%$. Without lensing, the CMB-SPA power spectra and the CMB distance prior give the same uncertainty. Correspondingly, relative to the CMB distance prior, the figure of merit (FoM) of the five knots, $[\det \mathbf{C}(X)]^{-1/2}$ \cite{Albrecht2006, Wang2008FoM}, is $13\%$ higher for the CMB-SPA power spectra with lensing and $2\%$ lower for those without lensing, as shown in \reffig{Xz_CMB_comparison_contour}. The CMB distance prior therefore retains essentially all the information of the primary CMB power spectra on $X(z)$, and is insensitive to the flat $\Lambda$CDM model assumed in the compression. To our knowledge, this is the first direct comparison between a CMB distance prior and the full CMB power-spectrum likelihood, including the lensing reconstruction, for a non-parametric reconstruction of the dark energy density.

\section{Projection of inter-probe tension into \texorpdfstring{$X(z)$}{X(z)}}
\label{sec:disc_Om_tension}

\begin{figure*}[tp]
    \centering
    \includegraphics[width=\linewidth]{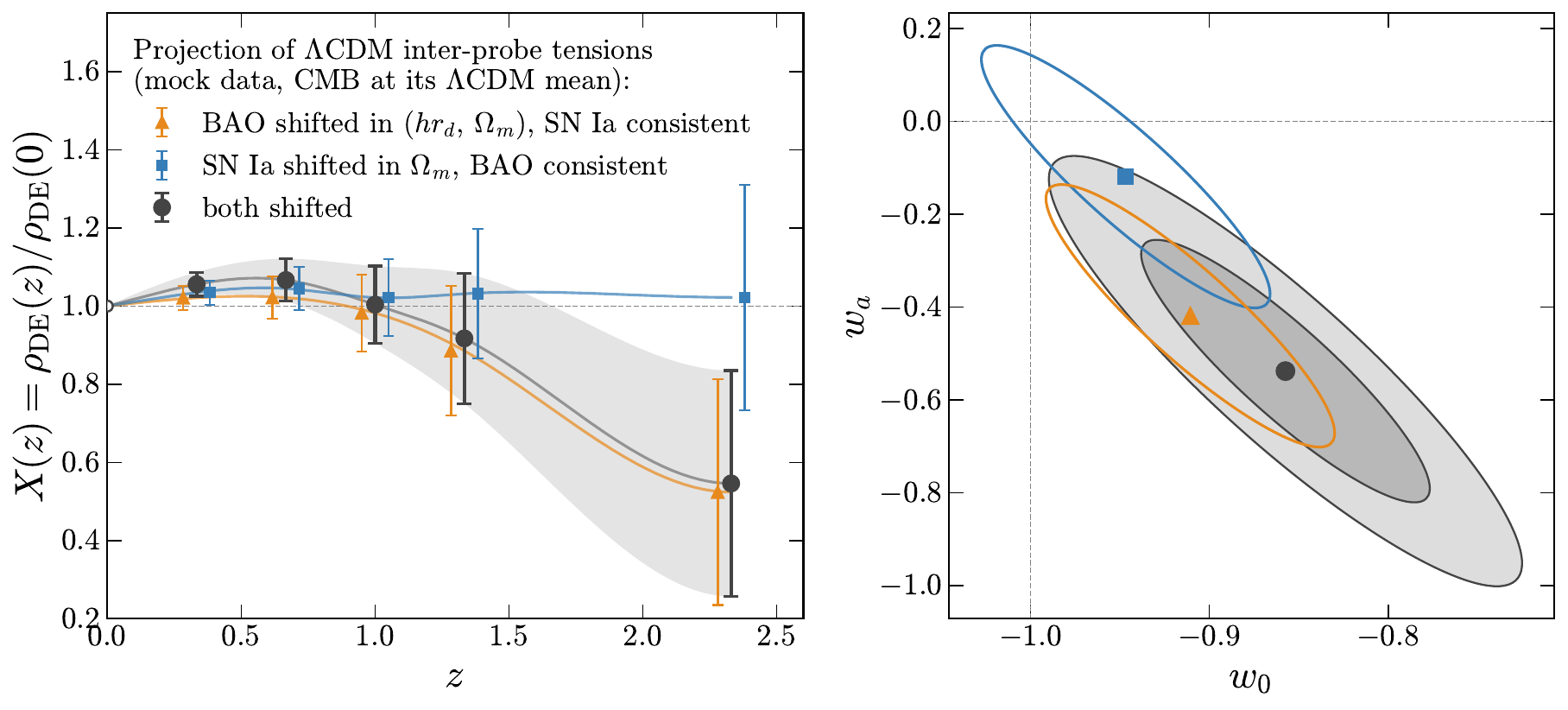}
    \caption{Projection of the $\Lambda$CDM tensions between CMB, BAO, and SNe~Ia in the $h r_d$--$\Omega_m$ plane into $X(z)$ (left) and $(w_0, w_a)$ (right), using the Fisher-matrix parameter shift of Equation~(\ref{eq:fisher_shift_main}). Noiseless data vectors are generated in flat $\Lambda$CDM at the CMB-SPA posterior mean, $(h r_d, \Omega_m) = (99.05\,{\rm Mpc}, 0.3154)$, with the BAO data shifted to the DESI~DR2 posterior mean, $(101.53\,{\rm Mpc}, 0.2975)$ (orange triangles), the SN~Ia data shifted to the Pantheon+ posterior mean, $\Omega_m = 0.3388$ (blue squares), or both (dark circles). \emph{Left:} Error bars show the $1\sigma$ uncertainties, offset horizontally for clarity. The gray curve and band show the combined shift of BAO and SN~Ia. \emph{Right:} The gray regions show the 68\% and 95\% C.L. ranges of the combined shift, and the orange and blue contours the 68\% C.L. ranges of the individual shifts. A $w_0w_a$CDM MCMC on the same noiseless data agrees with the Fisher shifts to within $0.09\sigma$. The dashed lines mark the cosmological constant.}
    \label{fig:Xz_fisher_decomposition}
    \vspace{12pt}
    \includegraphics[width=\linewidth]{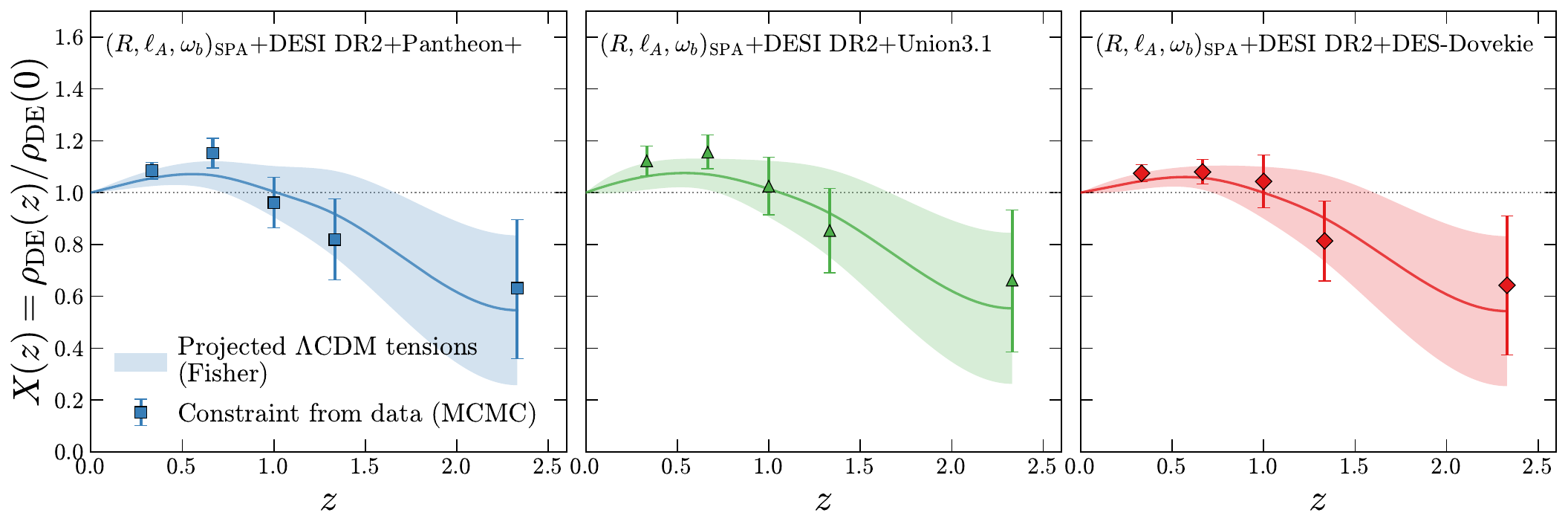}
    \caption{Comparison between the $X(z)$ predicted by projecting the inter-probe tensions under $\Lambda$CDM using Fisher matrix and the $X(z)$ reconstructed from the data. Each panel combines the CMB distance prior, DESI~DR2 BAO with one SN~Ia compilation: Pantheon+ (left), Union3.1 (middle), and DES-Dovekie (right). Solid curves and shaded bands show the predicted $X(z)$ and its $\pm 1\sigma$ ranges projected from their tensions in $h r_d$--$\Omega_m$ plane in $\Lambda$CDM. Data points with error bars show the $X(z)$ constrained from the real data and its 68\% C.L. ranges. For all three compilations, the constraints from the data are consistent with the projected $\Lambda$CDM tensions within the current uncertainties.}
    \label{fig:Xz_fisher_3panel}
\end{figure*}

The model-independent constraint of $X(z)$ gives the dark energy density evolution locally in redshift. This flexibility can provide a transparent view on how the tension between each pair of probes in the $\Lambda$CDM model projects into the dark energy density evolution $X(z)$, and can thus provide additional insight beyond the $(w_0,w_a)$ parameterization. Section~\ref{subsec:res_Xz} shows that $X(z)$ departs from unity in a see-saw shape. It lies above unity at $z \lesssim 2/3$ and falls below it at $z \gtrsim 4/3$. At the same time, the three probes disagree with each other under $\Lambda$CDM (Section~\ref{subsec:res_flatLCDM}). In this section we ask how much of the measured $X(z)$ these $\Lambda$CDM tensions alone can produce after being projected into $X(z)$ space. Because $X(z)$ is resolved in redshift, it shows separately how each tension projects onto the dark energy density evolution. We first project the {$h r_d$--$\Omega_m$} tension between BAO and the CMB, and the $\Omega_m$ tension between SNe~Ia and the CMB, into $X(z)$ with a Fisher-matrix calculation, and compare the result with the reconstruction from the data (Section~\ref{subsec:CMB-BAO-SN_tension}). We then examine the $H_0$ tension by including the SH0ES calibration of SNe~Ia (Section~\ref{subsec:CMB-SN_H0tension}).

\subsection{{$h r_d$--$\Omega_m$ tension} in $\Lambda$CDM model}
\label{subsec:CMB-BAO-SN_tension}
In $\Lambda$CDM, the tensions between the three probes can be visualized in the {$h r_d$--$\Omega_m$} plane (right panel of Figure~\ref{fig:Om_flatLCDM}). DESI~DR2 BAO is displaced from CMB-SPA along the CMB degeneracy direction, toward lower $\Omega_m$ and higher $h r_d$, at $2.5\sigma$, whereas the SNe~Ia constrain only $\Omega_m$ and prefer values higher than the CMB. To see how these displacements propagate into $X(z)$, we construct noiseless data vectors in flat $\Lambda$CDM. The distance prior $(R, \ell_A, \omega_b)$ is evaluated at the posterior mean of CMB-SPA, $({h r_d}, \Omega_m) = (99.05{\,{\rm Mpc}}, 0.3154)$ with $\omega_b = 0.02242$. The BAO data vector is evaluated at the $\Lambda$CDM posterior mean of DESI~DR2 alone, $({h r_d}, \Omega_m) = (101.53{\,{\rm Mpc}}, 0.2975)$, and the SN~Ia data vector at the $\Lambda$CDM posterior mean of $\Omega_m$ of each compilation. We assign to these data vectors the covariance matrices of the actual data and fit them with the same five-knot $X(z)$ model as in Section~\ref{subsec:res_Xz}. To first order, the best-fit shift of the parameters is (see Appendix~\ref{app:fisher} for the derivation)
\begin{equation}
    \Delta\boldsymbol{\theta} = \mathbf{F}^{-1}\sum_{p}\mathbf{J}_p^{\intercal}\mathbf{C}_p^{-1}\Delta\mathbf{d}_p\,, \qquad \mathbf{F} = \sum_{p}\mathbf{J}_p^{\intercal}\mathbf{C}_p^{-1}\mathbf{J}_p\,,
    \label{eq:fisher_shift_main}
\end{equation}
where $\boldsymbol{\theta} = (\omega_c, h, \omega_b, X_1, \ldots, X_5)$ are the parameters of the five-knot model with $X_k \equiv X(z_k)$, and $p$ labels the probe (CMB, BAO, and SNe Ia). For each probe, $\Delta\mathbf{d}_p$ is the shift of its data vector from the prediction of the CMB-SPA fiducial cosmology above ($\Delta\mathbf{d}_{\rm CMB} = 0$), $\mathbf{C}_p$ is its covariance matrix, and $\mathbf{J}_p \equiv \partial\mathbf{d}_{{\rm th},p}/\partial\boldsymbol{\theta}$ is the Jacobian of its theoretical prediction $\mathbf{d}_{{\rm th},p}$, evaluated at the fiducial cosmology with $X_k = 1$. The Fisher matrix $\mathbf{F}$ sums the information from all probes, and its inverse gives the parameter covariance. Table~\ref{tab:fisher_cov} in Appendix~\ref{app:fisher} lists the resulting knot values, marginalized uncertainties, and correlation coefficients of the five knots. The shift $\Delta\boldsymbol{\theta}$ is linear in the shifts of the data vectors, so the contributions from the BAO and the SN~Ia shifts are additive.

Figure~\ref{fig:Xz_fisher_decomposition} shows the $X(z)$ constraints from CMB+BAO+SNe~Ia datasets simulated in $\Lambda$CDM model but with realistic tension in the {$h r_d$--$\Omega_m$} plane. The BAO--CMB tension alone (orange triangles) produces the see-saw shape. This can be understood from the CMB. The CMB fixes $\omega_m$ and $r_d$, so the higher $h r_d$ preferred by BAO requires a higher $h$ and therefore a higher expansion rate at low redshift. To keep the angular diameter distance to the last scattering surface fixed by $\ell_A$, the expansion rate must then decrease at higher redshift, which, at fixed $\omega_m$, requires a lower dark energy density at $z \gtrsim 4/3$. The $\Omega_m$ tension between Pantheon+ and the CMB (blue squares) instead raises $X(z)$ almost uniformly, by $0.02$--$0.04$ at all knots, and thus produces a qualitatively different shape. Combining both (dark circles) gives $X = 1.056,\,1.067,\,1.004,\,0.917,\,0.546$ at the five knots. The largest deviations from unity, in units of the Fisher uncertainties, are $1.8\sigma$ at $z = 1/3$ and $1.6\sigma$ at $z = 2.33$, on the two sides of the see-saw. The see-saw is therefore mainly a projection of the BAO--CMB tension, to which the SN~Ia tension adds a small overall lift. The two tensions thus enter $X(z)$ in fundamentally different ways. The BAO--CMB tension involves the absolute distance scale $h r_d$, and since $\ell_A$ fixes the integrated distance to the last scattering surface, an excess of $X(z)$ at low redshift must be compensated by a deficit at high redshift. SNe~Ia constrain only relative distances, so the $\Omega_m$ tension between SNe~Ia and the CMB does not involve this constraint and is absorbed by a nearly uniform lift of $X(z)$ relative to $X(0) = 1$, without a sign change. To test the linear approximation directly, we also sample the $X(z)$ posterior with MCMC from the noiseless data vectors with the BAO and Pantheon+ shifts. The posterior means agree with the linear Fisher projection to within $0.12\sigma$ at all five knots, and the 68\% widths agree with the Fisher uncertainties to within $9\%$.

Figure~\ref{fig:Xz_fisher_3panel} compares the $X(z)$ predicted by tension projection between CMB, BAO, and SNe~Ia with the $X(z)$ jointly constrained from the actual data. Using the posterior covariance of the five knots, the measured $X(z)$ differs from the projected $X(z)$ by $\chi^2 = 3.09$, $2.20$, and $1.44$ for Pantheon+, Union3.1, and DES-Dovekie, respectively, compared with $\chi^2 = 15.7$, $12.5$, and $12.5$ relative to $X = 1$. This indicates that the measured $X(z)$ can be mostly reproduced by the projection of tension in $\Lambda$CDM model. The largest local deviation from the projection is $X(2/3)$ for Pantheon+, which lies $1.5\sigma$ above it. This is the same local excess identified in Section~\ref{subsec:res_Xz}, which cannot be fully absorbed by a projection of tension in $\Lambda$CDM but may point to another tension between datasets locally in redshift.

\subsection{Tension of Hubble constant}
\label{subsec:CMB-SN_H0tension}

\begin{figure}[t]
    \centering
    \includegraphics[width=\linewidth]{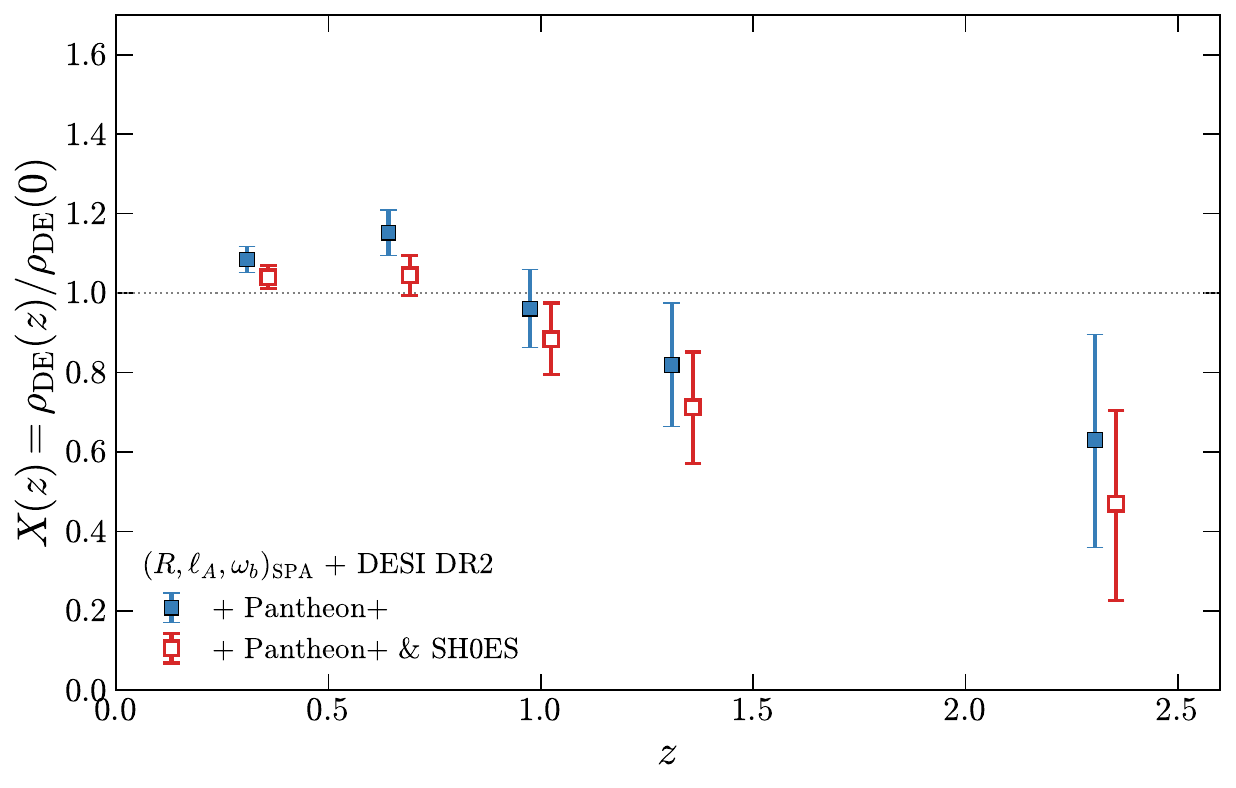}
    \caption{Constraints on $X(z)$ from {the CMB-SPA distance prior $(R,\ell_A,\omega_b)_{\rm SPA}$} + DESI DR2 + Pantheon+, with $M_B$ calibrated by SH0ES (open red squares) and with $M_B$ analytically marginalized over ({filled blue squares}). Error bars show the 68\% C.L. ranges at the five knots, and points at the same knot are offset horizontally for clarity. The dotted line marks the cosmological constant, $X = 1$.}
    \label{fig:Xz_SH0ES}
\end{figure}

\begin{figure}[h]
    \centering
    \includegraphics[width=\linewidth]{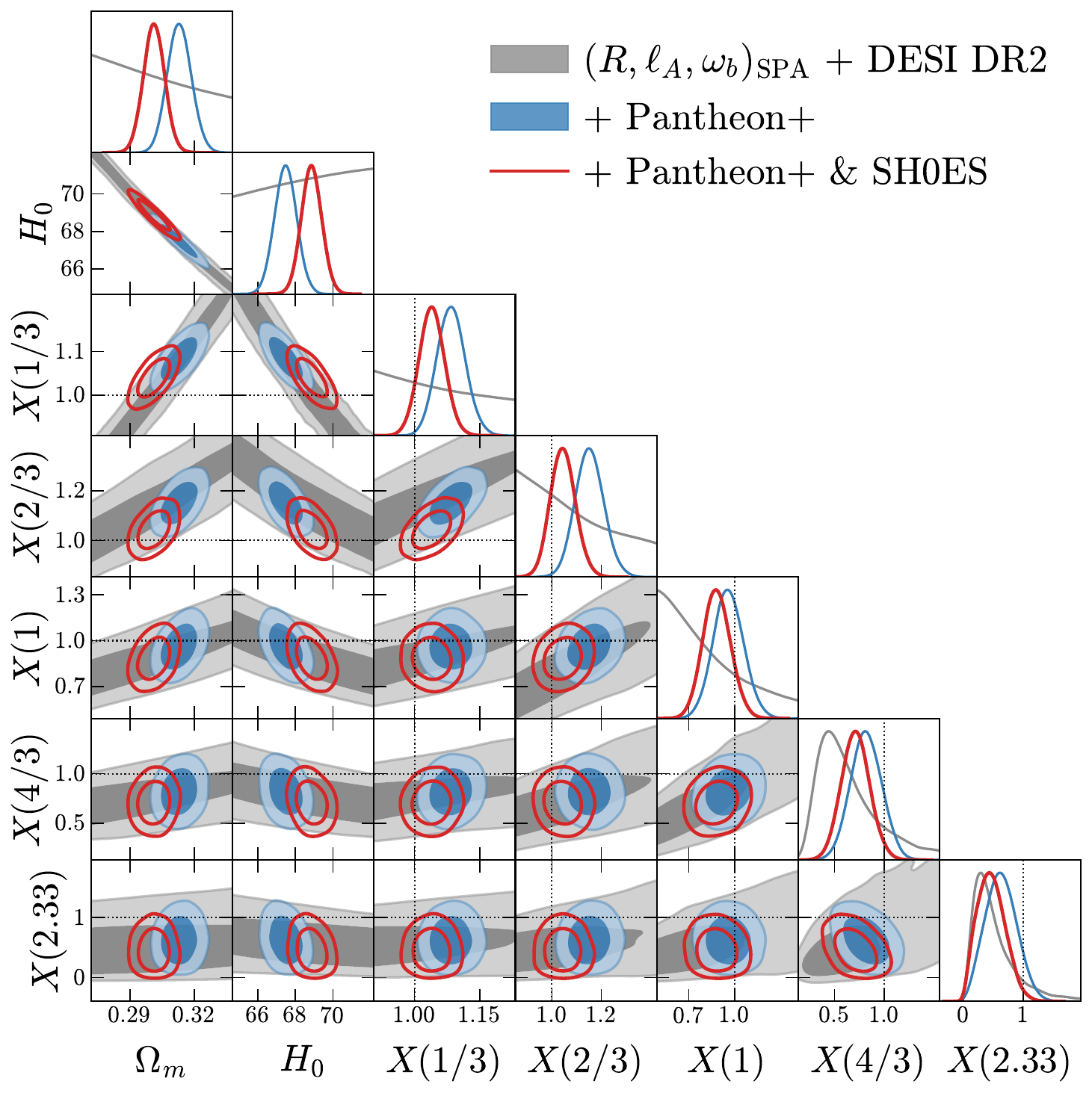}
    \caption{Marginalized posteriors of $\Omega_m$, $H_0$, and the dark energy density ratio at the five knots, $X(z_k)$, from the CMB-SPA distance prior, DESI~DR2 BAO, and Pantheon+, with $M_B$ analytically marginalized over (blue) and calibrated by SH0ES (red). Including the SH0ES calibration raises $H_0$, lowers $\Omega_m$, and shifts all $X(z_k)$ to lower values.}
    \label{fig:Xz_SH0ES_contour}
\end{figure}

The value of $H_0$ inferred from the CMB under $\Lambda$CDM model is different from the local distance-ladder measurement at $6.4\sigma$ for CMB-SPA and SH0ES \cite{Camphuis2026}. Measurements of $H_0$ independent of the sound horizon lie between the two, and differ from the distance ladder at $3.9\sigma$~\cite{Pantos2026}. So far we have analytically marginalized over the SN~Ia $H_0$ and $M_B$, so that SNe~Ia constrain only the shape of the distance--redshift relation and the Hubble tension does not enter the constraints of $X(z)$. Here we want to ask the opposite question: if the SN~Ia absolute calibration is included, how does the dark energy density $X(z)$ respond? We combine Pantheon+ with the SH0ES calibration \cite{Riess2022}, which constrains the absolute magnitude to $M_B = -19.253 \pm 0.027$ mag. 

Figure~\ref{fig:Xz_SH0ES} compares the $X(z)$ obtained from the joint CMB+BAO+Pantheon+ analysis with and without SH0ES calibration. Including SH0ES raises the inferred Hubble constant from $H_0 = 67.48^{+0.58}_{-0.59}$ to $68.86 \pm 0.54\,{\rm km\,s^{-1}\,Mpc^{-1}}$. Meanwhile, $X(z)$ decreases at all redshifts. The excess of $X(z)$ above unity at $z = 1/3$ and $2/3$ is now only $1.4\sigma$ and $0.9\sigma$, respectively. $X(z)$ at high redshifts declines, with $X(2.33)$ dropping from $0.63 \pm 0.27$ to $0.47\pm 0.24$. 

The response of $X(z)$ can be interpreted through the two-dimensional marginalized posteriors shown in Figure~\ref{fig:Xz_SH0ES_contour}. The CMB tightly constrains the physical matter density $\omega_m$, and its combination with BAO further constrains the physical dark energy density $\omega_{\rm DE}$. Since the SH0ES calibration introduces a preference for a higher Hubble constant, the joint analysis prefers a lower $\Omega_m$ than the one without SH0ES calibration and further shifts the $X(z)$ to lower values following the direction of degeneracy between $\Omega_m$ and $X(z_k)$ determined by CMB and BAO. A more intuitive way to understand this is through the normalization of $X(z)$. At low redshifts the Hubble parameter, which is fixed by BAO at $z > 0.295$ with the sound horizon $r_d$ calibrated by the CMB, can be computed as
\bea
\frac{H^2(z)}{[100\,{\rm km\,s^{-1}\,Mpc^{-1}}]^2} &{\simeq}& \omega_m(1+z)^3 + \omega_{\rm DE}(0)X(z),
\\
\omega_{\rm DE}(0) &=& h^2 - \omega_m {- \omega_\gamma}.
\eea
A preference for a higher $h$ from the SH0ES calibration therefore increases the present-day physical dark energy density $\omega_{\rm DE}(0)$ and hence decreases $X(z)$.

The results above show that the evolution of the dark energy density cannot fully absorb the $H_0$ tension. Including the SH0ES calibration would propagate the $H_0$ tension into $X(z)$. With the absolute expansion rate at $z \gtrsim 0.3$ fixed by BAO and $X(0) = 1$ by definition, $X(z)$ has little freedom to raise $H_0$. This result is expected because $H_0$ controls the global amplitude of distance $r(z)$ while the evolution of $X(z)$ alters the shape of $r(z)$.


\section{Discussion and Conclusions}
\label{sec:discussion}

We have measured the dark energy density ratio $X(z)\equiv\rho_{\rm DE}(z)/\rho_{\rm DE}(0)$ at five redshifts over $0<z<2.33$ from geometric probes, combining CMB-SPA, DESI DR2 BAO, and three updated SN~Ia compilations. For all three compilations, $X(z)$ lies above unity at $z\lesssim 2/3$, is consistent with unity at $z=1$, and falls below unity at $z\gtrsim 4/3$. Without flux averaging, the deviation from $X(z)=1$ is 2.3--2.8$\sigma$, compared with 3.0--3.2$\sigma$ in $w_0w_a$CDM. This lower significance mainly reflects the three additional degrees of freedom (Table~\ref{tab:dchi2}). With the updated compilations, the choice of SN~Ia sample changes the $w_0w_a$CDM significance by only 0.2$\sigma$, compared with the spread from 2.8$\sigma$ to 4.2$\sigma$ obtained with the original compilations~\cite{DESIDR2}. In $X(z)$CDM the spread increases to $0.5\sigma$, because with Pantheon+ the knot at $z=2/3$ rises above the $w_0w_a$CDM prediction, to $X(2/3)=1.15\pm0.06$, and improves the fit over $w_0w_a$CDM by $\Delta\chi^2_{\rm MAP}=4.65$, of which $3.0$ comes from the LRG2 BAO measurement at $z_{\rm eff}=0.706$. Out-of-sample predictive scoring likewise attributes the entire BAO preference for $w_0w_a$CDM over $\Lambda$CDM to LRG2~\cite{Karim2026}. With three more parameters than $w_0w_a$CDM, our additional improvement is not statistically significant by itself. However, the value of the model-independent approach goes beyond naive statistical arguments --- since we are ignorant of the nature of dark energy, it is critically important that we do not limit ourselves in probing it by using only simple parametric models which may not capture its true properties.

We have shown that the CMB distance prior reproduces the $X(z)$CDM posterior from the CMB power spectra without the lensing reconstruction likelihood, and that adding CMB lensing increases the information content only marginally. The CMB-SPA distance prior constructed in this paper can therefore be used as an external likelihood for future dark energy measurements, provided that pre-recombination physics, spatial flatness, and $\sum m_\nu=0.06$\,eV are as assumed in the compression. For distance priors beyond these assumptions, see~\cite{Pedrotti2026, ZhaiPark2020, CMBComp}.

Compared with the $(w_0, w_a)$ parameterization, $X(z)$ additionally shows that the tension between each pair of probes propagates into the dark energy density in fundamentally different ways. The see-saw shape of $X(z)$ primarily reflects the $h r_d$--$\Omega_m$ tension between CMB and BAO, while the $\Omega_m$ tension between SNe~Ia and the CMB appears as a slight, nearly flat excess of $X(z)$, and projecting these tensions in $\Lambda$CDM reproduces the measured evolution of $X(z)$. This picture agrees with the singular value decomposition of the BAO and SN~Ia distances in Ref.~\cite{Zaldarriaga2026}. In the space of the observables, the tension with the CMB concentrates in a single mode that measures $\Omega_m h^2$. Our analysis shows that $\{X(z_k)\}$ is a more informative basis than $(w_0, w_a)$ for understanding how the tensions project. It shows how each tension appears in the dark energy density when the probes are combined, and resolves features localized in redshift. How each tension projects onto the shape of $X(z)$ is set by the parameter dependency of each probe, and will therefore persist with the increasing precision of future CMB, BAO, and SN~Ia data if tension still exists. This highlights the importance of understanding the origin of the tension, if any, between BAO and CMB in the $\Lambda$CDM model, which drives the see-saw shape of the measured $X(z)$.

How should we interpret the detected evolution of $X(z)$? At this point, we cannot determine whether the result reflects genuine evolution of the dark energy density, a projection of the inter-probe tension due to unknown systematics, or a combination of both. If the dark energy density indeed followed a see-saw-shape evolution, we would observe the inter-probe $hr_d$--$\Omega_m$ tensions in $\Lambda$CDM and could not distinguish, with geometric probes alone, which one is the origin. One potential solution is to measure $X(z)$ from each pair of probes and check whether the results are consistent with each other. This solution is limited by the degeneracies left by each pair and by the current precision, but may become feasible with future BAO and SN~Ia data that are more precise, span more redshifts, and have better-understood systematics. An alternative method is to measure $\Omega_m$ within the $\Lambda$CDM model tomographically along the redshifts and to compare the probes at the same effective redshift~\cite{Colgain2022, Colgain2024c, Colgain2024a, Colgain2024b, ColgainSheikhJabbari2025, Colgain2026}. Since $E^2(z) \simeq \Omega_m(1+z)^3 + (1-\Omega_m)X(z)$ at low redshifts, the see-saw shape of $X(z)$ corresponds qualitatively to an effective $\Omega_m$ in $\Lambda$CDM that is higher than the CMB value at low redshift and lower at higher redshift. The tomographic $\Omega_m$ does not introduce dark energy parameters beyond the $\Lambda$CDM model and can be evaluated for each probe separately. Any inconsistency of $\Omega_m$ between probes at the same effective redshift could point to some unknown systematics.

One natural extension of our work is to perform the model-independent measurement of $X(z)$ with dynamical probes, such as $N$-point correlation function in 3D galaxy clustering \cite{Spezzati2026} and in weak lensing, assuming general relativity, and combine them with the compressed CMB and SN~Ia likelihoods constructed in this paper. The second extension is to measure $X(z)$ while freeing other parameters, such as the sum of neutrino masses~\cite{Craig2024, GreenMeyers2025, Elbers2025, Sailer2026, Ghedini2026, WangFreese2025}, spatial curvature~\cite{Chen2025, Zaldarriaga2026}, and the lensing amplitude $A_L$ or the optical depth $\tau$~\cite{Sailer2026, Jhaveri2025, Sullivan2026, Zaldarriaga2026, deCruzPerez2026, Park2025, PengPiao2025}, which requires recompressing the CMB distance prior. For the first extension, projecting the tension between geometric and dynamical probes in $\Lambda$CDM onto $X(z)$ as in Section~\ref{sec:disc_Om_tension} would show how it appears in the dark energy density evolution. An example is the lower $S_8$ reported by some weak lensing analyses~\cite{Abdalla2022, DESY3, DESY6}, although KiDS-Legacy finds $S_8$ consistent with the CMB~\cite{KiDSLegacy}. For the second one, the same method would show how a shift in each additional parameter projects onto $X(z)$, and whether it can mimic or absorb the see-saw shape. We leave these extensions to future work.

Future BAO and SN~Ia measurements will also test whether the $\Lambda$CDM parameters preferred by the individual probes converge as the measurement precision improves. The full five-year DESI survey~\cite{DESI2016} will test whether the BAO--CMB tension and the local feature at $z\simeq 2/3$ persist, and SPHEREx will add BAO measurements at low redshift from its all-sky galaxy survey~\cite{Dore2014, MataRoman2026}. At higher redshift, Euclid will measure BAO with H$\alpha$ emission-line galaxies over $0.9<z<1.8$~\cite{Laureijs2011, Blanchard2020, Mellier2025}, and the Roman High Latitude Wide Area Survey will map H$\alpha$ galaxies at $1<z<2$ and [OIII] galaxies at $2<z<3$~\cite{Spergel2015, Wang2022HLSS, ROTAC2025}, probing the redshift range where $X(z)$ crosses and falls below unity with tracers subject to different systematics, independently of the Ly$\alpha$ forest at $z\simeq 2.33$ \cite{DESIDR2, DESIDR2_LyaAP}. For SNe~Ia, the Rubin Observatory Legacy Survey of Space and Time (LSST) will provide a large sample at $z\lesssim1$~\cite{Ivezic2019, LSSTDESC2018}, and the Roman High Latitude Time Domain Survey is expected to deliver a cosmological sample of more than $10^4$ SNe~Ia~\cite{Hounsell2018, ROTAC2025, Rose2025}, about 40\% of them at $z>1$ in the reference survey design~\cite{Rose2021}. Because SNe~Ia measure relative distances independently of $r_d$, combining this high-redshift sample with the CMB can test whether the decline of $X(z)$ at high redshift, which our projection attributes to the $hr_d$ displacement of the BAO, persists without the BAO. These data will help determine whether the see-saw shape of $X(z)$ reflects the evolution of dark energy or the inter-probe tensions that currently reproduce it.
\\
\\
{\it AI usage.} This project was conceptualized, designed, and written by the authors. The first version of the pipeline was implemented by the authors in the \texttt{Julia} programming language, then converted to \texttt{Python} and further optimized with the large language model \texttt{Claude Opus 5}. The authors take full responsibility for the content of this paper.
\\
\\
{\it Code and data availability.} The CMB-SPA distance prior constructed in this paper are listed in Equation \ref{eq:cmbspa_prior}. The compressed SN~Ia likelihood for Pantheon+, Union3.1 and DES-Dovekie are tabulated in Appendix~\ref{app:logrp}. The pipeline code will be made publicly available upon the publication of this paper.

\begin{acknowledgments}
ZW would like to thank Fei Ge in particular for the detailed discussions on CMB-SPA likelihood and CMB lensing. ZW acknowledges Zhongxu Zhai, Jiachuan Xu, Nickolas Kokron, Seshadri Nadathur, Tanveer Karim, and Zhengyangguang (Laurence) Gong for helpful discussion. We also thank Eoin Ó Colgáin and Shahin Sheikh-Jabbari for useful comments on the first version of the manuscript.  We gratefully acknowledge support from NASA Grant \#80NSSC24M0021, ``Project Infrastructure for the Roman Galaxy Redshift Survey'', and NASA ROSES Grant 12-EUCLID11-0004.
\end{acknowledgments}

\appendix

\section{Fisher-matrix based parameter shift of tension in $\Lambda$CDM}
\label{app:fisher}

This appendix derives the Fisher-matrix based prediction for how inter-probe $hr_d$--$\Omega_m$ tension projects onto $X(z)$ knots.

The $X(z)$ reconstruction jointly fits 8 parameters $\boldsymbol{\theta} = (\omega_{c}, h, \omega_b, X_1, X_2, X_3, X_4, X_5)$, with $\sum m_\nu = 0.06$\,eV fixed, to three data vectors: the CMB distance prior $(R, \ell_A, \omega_b)$ (Equation~\eqref{eq:cmbspa_prior}), DESI DR2 BAO ($D_M/r_d$ and $D_H/r_d$ at 6 redshifts, plus the BGS $D_V/r_d$ measurement), and the compressed SN~Ia likelihood, i.e. $\log r_p$ at 11 redshifts (Section~\ref{sec:SNe_flux_avg}), in which $\mathcal{M}$ is already marginalized over.

The Fisher information matrix is defined as the expectation of the negative Hessian of the log-likelihood,
\begin{equation}
    F_{\alpha\beta} \equiv -\left\langle \frac{\partial^2 \log\mathcal{L}}{\partial \theta_\alpha\, \partial \theta_\beta} \right\rangle\,.
    \label{eq:fisher_def}
\end{equation}
We assume a Gaussian likelihood
\begin{equation}
    -2\log\mathcal{L} = [\mathbf{d} - \mathbf{d}_{\rm th}(\boldsymbol{\theta})]^{\intercal} \mathbf{C}^{-1} [\mathbf{d} - \mathbf{d}_{\rm th}(\boldsymbol{\theta})] + {\rm const.}\,,
\end{equation}
where $\mathbf{d}$ is the observed data vector, $\mathbf{d}_{\rm th}(\boldsymbol{\theta})$ is the theoretical prediction, and $\mathbf{C}$ is the data covariance matrix which is independent of the parameters. Differentiating twice and taking the expectation (noting that $\langle \mathbf{d} - \mathbf{d}_{\rm th}\rangle = 0$ so the second-derivative term of $\mathbf{d}_{\rm th}$ vanishes) gives
\begin{equation}
     F_{\alpha\beta} = \frac{\partial \mathbf{d}_{\rm th}^{\intercal}}{\partial \theta_\alpha}\, \mathbf{C}^{-1}\, \frac{\partial \mathbf{d}_{\rm th}}{\partial \theta_\beta} = \left(\mathbf{J}^{\intercal} \mathbf{C}^{-1} \mathbf{J}\right)_{\alpha\beta}\,,
\end{equation}
where $J_{i\alpha} \equiv \partial d_{{\rm th},i}/\partial \theta_\alpha$ is the Jacobian of the theoretical data vector with respect to the model parameters. Since the three probes are independent, the total Fisher matrix is the sum of the individual contributions:
\begin{equation}
    \begin{aligned}
    F_{\alpha\beta} = {}& J_{\rm CMB}^{\intercal} C_{\rm CMB}^{-1} J_{\rm CMB} + J_{\rm BAO}^{\intercal} C_{\rm BAO}^{-1} J_{\rm BAO} \\
    & + J_{\rm SNe}^{\intercal} C_{\rm SNe}^{-1} J_{\rm SNe}\,,
    \end{aligned}
    \label{eq:fisher}
\end{equation}
where the Jacobians are evaluated numerically at the flat $\Lambda$CDM posterior mean of CMB-SPA, $(\Omega_m, h, \omega_b) = (0.3154, 0.6735, 0.02242)$ and $X_k = 1$. $C_{\rm SNe}$ is the covariance of the compressed $\log r_p$, so that no further projection of the offset $\mathcal{M}$ is needed.

We construct noise-free data vectors from theory. In the fiducial case, all three probes (CMB, BAO, and SNe~Ia) share the same $\Lambda$CDM cosmology above, with $X(z) \equiv 1$, and no inter-probe tension. Since the data vectors are generated directly from the fiducial theory without noise, $\chi^2 = 0$ by construction. We then ask: what happens when a probe's underlying cosmology differs from the fiducial? This simulates the inter-probe tension observed in real data.

The CMB data vector is kept at the fiducial. The BAO data vector is shifted to flat $\Lambda$CDM at the posterior mean of DESI~DR2 alone, $(h r_d, \Omega_m) = (101.53\,{\rm Mpc}, 0.2975)$, at fixed $\omega_b$, with $h$ solved for from $h r_d$. To separate the two components of this shift, we also evaluate it at $(h r_d, \Omega_m) = (99.05\,{\rm Mpc}, 0.2975)$ ($\Omega_m$ only) and at $(101.53\,{\rm Mpc}, 0.3154)$ ($h r_d$ only). The SN~Ia data vector depends only on $\Omega_m$ and is shifted to the posterior mean of $\Omega_m$ of each compilation in flat $\Lambda$CDM (Table~\ref{tab:CPL}). The shift of each data vector is $\Delta\mathbf{d}_p = \mathbf{d}_{{\rm th},p}(\boldsymbol{\theta}_p) - \mathbf{d}_{{\rm th},p}(\boldsymbol{\theta}_0)$, where $p$ labels the probe (CMB, BAO, SN~Ia). To find the best-fit parameter shift, we minimize $\chi^2$ with the shifted data. Since the fiducial data vector is $\mathbf{d}_{0} \equiv \mathbf{d}_{\rm th}(\boldsymbol{\theta}_0)$, the residual of probe $p$ at $\boldsymbol{\theta}_0 + \Delta\boldsymbol{\theta}$ is $\mathbf{d}_{0,p} + \Delta\mathbf{d}_p - \mathbf{d}_{{\rm th},p}(\boldsymbol{\theta}_0 + \Delta\boldsymbol{\theta}) = \Delta\mathbf{d}_p - \mathbf{J}_p\,\Delta\boldsymbol{\theta}$ to first order, and the total $\chi^2$ becomes
\begin{equation}
    \chi^2 \approx \sum_{p} \left(\Delta\mathbf{d}_p - \mathbf{J}_p\,\Delta\boldsymbol{\theta}\right)^{\intercal} \mathbf{C}_p^{-1} \left(\Delta\mathbf{d}_p - \mathbf{J}_p\,\Delta\boldsymbol{\theta}\right).
\end{equation}
Setting $\partial\chi^2/\partial\Delta\theta_\alpha = 0$ gives
\begin{equation}
    \sum_{p} \mathbf{J}_p^{\intercal} \mathbf{C}_p^{-1} \mathbf{J}_p\,\Delta\boldsymbol{\theta} = \sum_{p} \mathbf{J}_p^{\intercal} \mathbf{C}_p^{-1} \Delta\mathbf{d}_p\,,
\end{equation}
i.e. $\mathbf{F}\,\Delta\boldsymbol{\theta} = \sum_{p} \mathbf{J}_p^{\intercal} \mathbf{C}_p^{-1} \Delta\mathbf{d}_p$. The best-fit parameter shift is
\begin{equation}
    \Delta\boldsymbol{\theta} = \mathbf{F}^{-1} \sum_{p} \mathbf{J}_p^{\intercal} \mathbf{C}_p^{-1} \Delta\mathbf{d}_p\,.
    \label{eq:fisher_shift}
\end{equation}
The predicted shift in each $X(z)$ knot is $\Delta X_k = \Delta\theta_{3+k}$ ($k = 1, \ldots, 5$).

\begin{table*}[t]
\centering
\caption{Fisher-matrix projection of the five $X(z)$ knots from $(R,\ell_A,\omega_b)_{\rm SPA}$ + DESI DR2 + SN~Ia. The $X \pm \sigma$ rows give the knot values projected from the BAO and SN~Ia shifts, $X_k = 1 + \Delta X_k$, with the uncertainties $\sigma_{X_k}$. The $\rho$ rows give the correlation coefficients between the knot in the subscript of $\rho$ and the knot of each column. The uncertainties and correlations are marginalized over $\omega_c$, $h$, and $\omega_b$, at the flat $\Lambda$CDM fiducial of CMB-SPA.}
\label{tab:fisher_cov}
\small
\renewcommand{\arraystretch}{1.15}
\setlength{\tabcolsep}{5pt}
\begin{tabular}{@{}llccccc@{}}
\toprule
SN~Ia & & $X(1/3)$ & $X(2/3)$ & $X(1)$ & $X(4/3)$ & $X(2.33)$ \\
\midrule
Pantheon+ & $X \pm \sigma$ & $1.056 \pm 0.031$ & $1.067 \pm 0.055$ & $1.004 \pm 0.099$ & $0.917 \pm 0.166$ & $0.546 \pm 0.289$ \\
\cmidrule{2-7}
 & $\rho_{X(2/3)}$ & $0.51$ &  &  &  &  \\
 & $\rho_{X(1)}$ & $0.00$ & $0.25$ &  &  &  \\
 & $\rho_{X(4/3)}$ & $0.12$ & $-0.02$ & $0.19$ &  &  \\
 & $\rho_{X(2.33)}$ & $0.05$ & $0.07$ & $-0.05$ & $-0.39$ &  \\
\midrule
Union3.1 & $X \pm \sigma$ & $1.063 \pm 0.052$ & $1.070 \pm 0.060$ & $1.014 \pm 0.108$ & $0.920 \pm 0.169$ & $0.554 \pm 0.291$ \\
\cmidrule{2-7}
 & $\rho_{X(2/3)}$ & $0.55$ &  &  &  &  \\
 & $\rho_{X(1)}$ & $0.34$ & $0.33$ &  &  &  \\
 & $\rho_{X(4/3)}$ & $0.21$ & $0.05$ & $0.24$ &  &  \\
 & $\rho_{X(2.33)}$ & $0.13$ & $0.10$ & $0.01$ & $-0.36$ &  \\
\midrule
DES-Dovekie & $X \pm \sigma$ & $1.045 \pm 0.031$ & $1.057 \pm 0.046$ & $1.000 \pm 0.100$ & $0.901 \pm 0.167$ & $0.543 \pm 0.289$ \\
\cmidrule{2-7}
 & $\rho_{X(2/3)}$ & $0.36$ &  &  &  &  \\
 & $\rho_{X(1)}$ & $0.10$ & $0.27$ &  &  &  \\
 & $\rho_{X(4/3)}$ & $0.12$ & $-0.05$ & $0.20$ &  &  \\
 & $\rho_{X(2.33)}$ & $0.06$ & $0.06$ & $-0.04$ & $-0.38$ &  \\
\bottomrule
\end{tabular}
\end{table*}

The marginalized uncertainties and covariance of the knots are given by the $5\times5$ block of $\mathbf{F}^{-1}$ for $(X_1,\ldots,X_5)$, which is marginalized over $\omega_c$, $h$, and $\omega_b$. Table~\ref{tab:fisher_cov} lists the uncertainties $\sigma_{X_k} = [(\mathbf{F}^{-1})_{3+k,\,3+k}]^{1/2}$ and the correlation coefficients $\rho_{kl} = (\mathbf{F}^{-1})_{3+k,\,3+l}/(\sigma_{X_k}\sigma_{X_l})$ for each SN~Ia compilation. Since the shifts in Equation~(\ref{eq:fisher_shift}) are evaluated with the same $\mathbf{F}$, these uncertainties do not depend on the tension that is projected.

\section{Compressed SN~Ia likelihoods}
\label{app:logrp}
For reproducibility, we list the compressed SN~Ia likelihoods used in this paper. Table~\ref{tab:logrp} gives the data vectors $\boldsymbol y$ of the 11 $\log r_p$ knots (Equation~\eqref{eq:rp}) and their covariance matrices $\boldsymbol\Sigma_y$, in which $\mathcal{M}$ has been analytically marginalized over, for Pantheon+, Union3.1, and DES-Dovekie without flux averaging. The theoretical prediction is $y^{\rm th}_k = \log r_p(z_k)$, computed from the comoving distance of each model with Equation~(\ref{eq:rp}). Because the knots are strongly correlated, with correlation coefficients up to $0.99$ for Union3.1, we quote $\boldsymbol\Sigma_y$ to five significant figures.

\begin{table*}[p]
\centering
\caption{Compressed SN~Ia likelihoods of the updated Pantheon+, Union3.1 (UNITY1.7), and DES-Dovekie compilations. $y_k = \log r_p(z_k)$ is the mean value at the 11 redshift knots $\{z_k\}$, and the covariance matrices $\boldsymbol\Sigma_y$ are in units of $10^{-4}$. The DES-Dovekie values are the same as those without flux averaging in Table~VII of Paper~II~\cite{Wang2026SNcompression}, whereas the Pantheon+ and Union3.1 values are from the updated SN~Ia data.}
\label{tab:logrp}
\renewcommand{\arraystretch}{1.15}
\setlength{\tabcolsep}{4.5pt}
\resizebox{\textwidth}{!}{%
\begin{tabular}{lrrrrrrrrrrrr}
\toprule
$z_k$ & \multicolumn{1}{c}{$y_k$} & \multicolumn{11}{c}{$\Sigma_{y,kl}\ [10^{-4}]$} \\
\midrule
\multicolumn{13}{l}{\textbf{Pantheon+}} \\[3pt]
$0.05$ & $-0.00321$ & $1.3581$ &  &  &  &  &  &  &  &  &  &  \\
$0.15$ & $-0.01165$ & $1.1367$ & $1.2771$ &  &  &  &  &  &  &  &  &  \\
$0.3$ & $-0.00816$ & $1.2476$ & $1.1498$ & $1.4776$ &  &  &  &  &  &  &  &  \\
$0.4$ & $-0.01322$ & $1.1128$ & $1.1442$ & $1.1996$ & $1.4907$ &  &  &  &  &  &  &  \\
$0.5$ & $-0.02568$ & $1.2427$ & $1.1547$ & $1.3373$ & $1.2212$ & $1.9959$ &  &  &  &  &  &  \\
$0.6$ & $-0.02525$ & $1.1756$ & $1.1613$ & $1.2669$ & $1.2642$ & $1.2792$ & $2.2892$ &  &  &  &  &  \\
$0.7$ & $-0.02471$ & $1.2069$ & $1.1545$ & $1.3127$ & $1.2443$ & $1.4679$ & $1.3799$ & $2.9487$ &  &  &  &  \\
$0.8$ & $-0.04396$ & $1.1702$ & $1.1491$ & $1.2936$ & $1.2615$ & $1.3095$ & $1.4820$ & $0.93040$ & $4.8747$ &  &  &  \\
$1$ & $0.05259$ & $1.1250$ & $1.1112$ & $1.2694$ & $1.1409$ & $1.1616$ & $1.2153$ & $0.59924$ & $2.9942$ & $8.4931$ &  &  \\
$1.6$ & $-0.04617$ & $1.1477$ & $1.1203$ & $1.2903$ & $1.1748$ & $1.1917$ & $1.2473$ & $0.66856$ & $2.4965$ & $-0.41121$ & $16.374$ &  \\
$2.27$ & $0.00422$ & $1.2018$ & $1.1374$ & $1.4197$ & $1.2878$ & $1.4569$ & $1.2071$ & $1.5616$ & $0.96829$ & $6.0037$ & $-0.18320$ & $72.455$ \\
\midrule
\multicolumn{13}{l}{\textbf{Union3.1 (UNITY1.7)}} \\[3pt]
$0.05$ & $0.02982$ & $14.458$ &  &  &  &  &  &  &  &  &  &  \\
$0.15$ & $0.02722$ & $16.894$ & $20.471$ &  &  &  &  &  &  &  &  &  \\
$0.3$ & $0.01270$ & $17.174$ & $20.566$ & $21.176$ &  &  &  &  &  &  &  &  \\
$0.4$ & $0.02183$ & $16.381$ & $19.842$ & $20.190$ & $19.756$ &  &  &  &  &  &  &  \\
$0.5$ & $0.01512$ & $16.969$ & $20.395$ & $20.841$ & $20.041$ & $21.204$ &  &  &  &  &  &  \\
$0.6$ & $0.02077$ & $16.602$ & $20.048$ & $20.466$ & $19.760$ & $20.362$ & $20.750$ &  &  &  &  &  \\
$0.7$ & $-0.00657$ & $16.758$ & $20.189$ & $20.678$ & $19.884$ & $20.629$ & $20.367$ & $21.791$ &  &  &  &  \\
$0.8$ & $-0.00261$ & $16.711$ & $20.147$ & $20.627$ & $19.871$ & $20.600$ & $20.411$ & $20.907$ & $21.863$ &  &  &  \\
$1$ & $0.03630$ & $16.746$ & $20.193$ & $20.645$ & $19.903$ & $20.684$ & $20.348$ & $20.713$ & $20.727$ & $23.670$ &  &  \\
$1.6$ & $-0.02979$ & $16.621$ & $20.032$ & $20.505$ & $19.795$ & $20.404$ & $20.329$ & $20.115$ & $21.419$ & $18.558$ & $39.814$ &  \\
$2.27$ & $-0.07128$ & $16.981$ & $20.475$ & $21.017$ & $20.255$ & $21.169$ & $20.547$ & $21.740$ & $19.854$ & $25.689$ & $13.964$ & $251.33$ \\
\midrule
\multicolumn{13}{l}{\textbf{DES-Dovekie}} \\[3pt]
$0.05$ & $-0.00598$ & $3.3977$ &  &  &  &  &  &  &  &  &  &  \\
$0.15$ & $-0.02372$ & $3.0878$ & $3.5860$ &  &  &  &  &  &  &  &  &  \\
$0.3$ & $-0.01897$ & $3.1574$ & $2.9429$ & $3.4040$ &  &  &  &  &  &  &  &  \\
$0.4$ & $-0.02190$ & $2.9146$ & $2.9617$ & $2.9457$ & $3.0960$ &  &  &  &  &  &  &  \\
$0.5$ & $-0.02182$ & $3.0704$ & $2.9455$ & $3.1391$ & $2.9108$ & $3.3389$ &  &  &  &  &  &  \\
$0.6$ & $-0.01900$ & $3.0141$ & $2.9771$ & $3.0411$ & $2.9112$ & $2.9747$ & $3.2364$ &  &  &  &  &  \\
$0.7$ & $-0.01238$ & $3.0873$ & $2.9768$ & $3.1118$ & $2.8903$ & $3.1170$ & $2.9770$ & $3.4865$ &  &  &  &  \\
$0.8$ & $-0.02204$ & $3.0368$ & $2.9976$ & $3.0776$ & $2.9158$ & $3.0181$ & $3.0676$ & $3.0688$ & $3.7874$ &  &  &  \\
$0.9$ & $-0.02615$ & $3.0197$ & $2.9535$ & $3.1152$ & $2.9425$ & $3.1318$ & $3.0129$ & $3.1549$ & $3.0522$ & $5.3638$ &  &  \\
$1$ & $-0.03581$ & $3.0666$ & $3.0041$ & $3.0858$ & $2.9138$ & $3.0667$ & $3.0840$ & $3.1501$ & $3.1945$ & $3.0621$ & $5.9642$ &  \\
$1.15$ & $-0.06860$ & $2.9687$ & $2.9349$ & $3.1419$ & $2.9692$ & $3.1757$ & $3.0394$ & $3.1610$ & $3.0861$ & $3.9734$ & $1.3943$ & $21.726$ \\
\bottomrule
\end{tabular}}
\end{table*}

\bibliography{main}

\end{document}